# X-Ray Spectral Variability as a Probe of Multimessenger Emission in Blazar 5BZB J0630−2406

Jose Maria Sanchez Zaballa[1], Sara Buson[1,2], Stefano Marchesi[3,4,5], Francesco Tombesi[6,7,8], Thomas Dauser[9], Joern Wilms[9], and Alessandra Azzollini[1]

[1] Julius-Maximilians-Universität Würzburg, Fakultät für Physik und Astronomie, Institut für Theoretische Physik und Astrophysik, Lehrstuhl für Astronomie, Emil-Fischer-Str. 31, D-97074 Würzburg, Germany; jose.sanchez-zaballa@uni-wuerzburg.de
[2] Deutsches Elektronen-Synchrotron DESY, Platanenallee 6, 15738 Zeuthen, Germany
[3] Dipartimento di Fisica e Astronomia (DIFA) Augusto Righi, Università di Bologna, via Gobetti 93/2, I-40129 Bologna, Italy
[4] Department of Physics and Astronomy, Clemson University, Kinard Lab of Physics, Clemson, SC 29634, USA
[5] INAF—Osservatorio di Astrofisica e Scienza dello Spazio (OAS), via Gobetti 93/3, I-40129 Bologna, Italy
[6] Physics Department, Tor Vergata University of Rome, Via della Ricerca Scientifica 1, 00133 Rome, Italy
[7] INAF—Astronomical Observatory of Rome, Via Frascati 33, 00040 Monte Porzio Catone, Italy
[8] INFN—Rome Tor Vergata, Via della Ricerca Scientifica 1, 00133 Rome, Italy
[9] Dr. Remeis-Sternwarte and ECAP, Friedrich-Alexander-Universität Erlangen-Nürnberg, Sternwartstr. 7, 96049 Bamberg, Germany

## Abstract

X-ray observations are essential for understanding the multimessenger emission mechanisms of active galactic nuclei (AGN). Blazars, a subset of AGN whose X-ray emission predominantly originates from relativistic jets, have been proposed as promising high-energy neutrino sources. In this work, we study the candidate neutrino-emitting blazar 5BZB J0630-2406, which has been observed over multiple epochs with the XMM-Newton, NuSTAR, Neil Gehrels Swift-XRT, and eROSITA observatories. Analysis of the X-ray spectra in the 2.0–10.0 keV band shows significant variability, with high-flux states adhering to a power-law model indicative of jet emission. However, during low-flux states, the spectrum reveals an additional component in hard X-rays, indicating a transition from jet-dominated to multicomponent X-ray emission, possibly associated with hadronic processes. To investigate this spectral evolution, we tested various models and found it to be consistent with coronal emission or photoionized absorption processes typically observed in obscured AGN. The identification of the X-ray spectral variability in 5BZB J0630-2406, combined with its potential for neutrino production, opens new perspectives in multimessenger astrophysics of blazars, highlighting the synergies between the mechanisms of the jet and the nuclear environment.

## 1. Introduction

Active galactic nuclei (AGN) are prime targets for multiwavelength observations due to their variable emission, which spans the entire electromagnetic spectrum, from radio wavelengths to $\gamma$-rays. Among them, blazars host powerful relativistic jets pointing toward Earth, which boost their emissions. The emission processes in blazars can be explained by leptonic and lepto-hadronic models. In leptonic models, the spectral energy distribution (SED) of blazars is primarily explained by synchrotron radiation and inverse Compton scattering by relativistic electrons. Lepto-hadronic models suggest that hadrons are accelerated along with leptons, leading to processes such as pion production and synchrotron radiation from secondary particles (K. Mannheim 1993). A natural product of these secondary interactions are neutrinos. In lepto-hadronic scenarios, neutrinos may originate in an optically thick environment where the suppression of $\gamma$-ray emission at the highest energies ($E \sim$ GeV–TeV) due to self-absorption could lead to an enhanced flux in the soft-to-hard X-ray range (M. Petropoulou et al. 2015, 2020; K. Murase et al. 2016; A. Reimer et al. 2019; F. Oikonomou et al. 2021). Observations in X-rays provide crucial insights into the underlying physical processes, while also having the potential to distinguish between leptonic and hadronic emission scenarios (e.g., H. Zhang et al. 2019).

A case study highlighting the potential distinction between scenarios is presented by the blazar 5BZB J0630-2406, where previous attempts to model the observed electromagnetic emission using conventional leptonic models have encountered challenges (M. Ackermann et al. 2016). The object recently gained further interest following the proposed association with an IceCube neutrino hotspot, over 2008–2015, and inclusion among the sample of candidate "PeVatron blazars" (S. Buson et al. 2022a, 2022b, 2023, the last hereafter Paper I). With a redshift constraint of $1.239 < z < 1.33$ (M. S. Shaw et al. 2013; M. Lainez et al. 2023), it has been historically classified as a BL Lacertae (BL Lac) object due to its featureless optical spectrum and high synchrotron peak, with $\nu_{sy} \sim 10^{15}$ Hz. These blazars generally lack strong radiation fields, and their SED is well described by a simple one-zone leptonic model, i.e., a synchrotron self-Compton model. However, the analysis of its quasi-simultaneous multiwavelength SED revealed that 5BZB J0630-2406 is intrinsically a "high-power blue flat spectrum radio quasar" (G. Ghisellini et al. 2012), also know as a "masquerading BL Lac" (P. Padovani et al. 2019). It hosts a standard accretion disk and broad-line region (BLR), including a powerful jet and radiatively efficient accretion ($L_\gamma/L_{\rm Edd} \sim 1.02$, $L_{\rm BLR}/L_{\rm Edd} < 5.79 \times 10^{-4}$, A. Azzollini et al. 2025, hereafter Paper PI) with similarities to other candidate neutrino-emitter blazars such as TXS 0506+056, the first high-energy neutrino source detected by IceCube (IceCube Collaboration et al. 2018; P. Padovani et al. 2019).

**Table 1**
Summary of Observational Epochs with Corresponding Modified Julian Date (MJD), Calendar Date, Exposure Time (Expo.), and Instruments Used

| Epoch | MJD | Date | Expo. (ks) | ObsID | Instruments |
|---|---|---|---|---|---|
| T1 (a) | 54867 | 2009-02-05 | 5.3 | 00038384 | Swift-XRT |
| T1 (b) | 55382 | 2010-07-05 | 4.0 | 00040857 | Swift-XRT |
| T1 (c) | 55541 | 2010-12-11 | 1.3 | 00041690001 | Swift-XRT |
| T2 (a) | 56948 | 2014-10-18 | 2.3 | 00080776001 | Swift-XRT |
| | | | 9.0 | 0740820401 | XMM-Newton |
| | | | 66.6 | 60001140002 | NuSTAR |
| T2 (b) | 56970 | 2014-11-09 | 5.0 | 00091900001 | Swift-XRT |
| T3 | 58948 | 2020-04-09 | 0.2 | 1eRASS J063136.5-240950 | eROSITA |
| T4 (a) | 60477 | 2024-06-01 | 3.2 | 00041690002 | Swift-XRT |
| T4 (b) | 60462 | 2024-06-05 | 1.7 | 00041690003 | Swift-XRT |
| T4 (c) | 60476 | 2024-06-15 | 3.5 | 00097520002 | Swift-XRT |
| | | | 18.9 | 61060004001 | NuSTAR |
| T4 (d) | 60490 | 2024-06-29 | 5.0 | 00041690004 | Swift-XRT |

Previous attempts to model the SED showed that a pure self-synchrotron model, which is generally used for BL Lac objects, faces challenges in adequately reproducing the broadband SED (M. Ackermann et al. 2016). In an attempt to improve the fit, external-Compton models, which are typically applied to flat-spectrum radio quasars, were also tested but failed to provide a satisfactory representation of the SED, with the X-ray data displaying the largest deviation from the model. The authors found that the only option to model the SED with conventional leptonic models was to remove the X-ray points from the fits and assign them to an additional, unmodeled component. In our previous work, the SED has been modeled using both leptonic and lepto-hadronic scenarios, suggesting that the hadronic component is subdominant except in the X-ray and MeV bands (G. Fichet de Clairfontaine et al. 2023, hereafter Paper TI). The analysis of the simultaneous XMM-Newton and NuSTAR spectra provided evidence (at $\gtrsim 3\sigma$) of a break in the X-ray band, which is challenging to reproduce by purely leptonic models. On the other hand, the break in the X-ray band, assuming lepto-hadronic models, could be interpreted as secondaries from the hadronic cascade, i.e., synchrotron emission from leptonic pairs generated via Bethe–Heitler pair production.

Motivated by the observed break in the X-ray spectrum, and its multimessenger implications (e.g., neutrino production), we systematically studied archival X-ray observations along with newly granted Swift-XRT/NuSTAR observations of 5BZB J0630-2406, to further investigate the nature of the X-ray emission in this blazar.

This paper is structured as follows: in Section 2, we describe the data sets used for our analysis, including the specific reduction procedures for each instrument. In Section 3, we present a detailed investigation of the X-ray spectral properties, testing both simple power-law and more complex models across all epochs. We then address the consistency of outcomes across different epochs with tests performed in Section 4. The observed X-ray spectral and variability properties are summarized in Section 5, while their physical origin is tested in Section 6. Then we put the findings into the broader context of neutrino multimessenger astrophysics in Section 7. Finally, multimessenger implications and conclusions are presented in Section 8.

## 2. X-Ray Data Analysis

In this section, we present the X-ray observations and analysis of data available from Swift-XRT, XMM-Newton, NuSTAR, and eROSITA. Observations available for four different epochs are referenced throughout the paper as summarized in Table 1.

We anticipate that epoch T2 (a) offers the highest-quality data set from the statistical point of view, encompassing simultaneous observations performed with XMM-Newton, NuSTAR, and Swift-XRT, which were analyzed in our previous study (Paper TI). Additionally, epoch T4 includes a joint observation with NuSTAR and Swift-XRT, while the remaining observations encompass data from Swift-XRT or eROSITA. The light curve displaying intrinsic fluxes for all epochs is presented in Figure 1.

### 2.1. Swift-XRT

Swift-XRT data (N. Gehrels et al. 2004) were processed using FTOOLS (v0.13.7) within the HEASOFT package (v6.32) for analyzing FITS files. For Swift observations, the Swift-XRT data were collected in photon counting mode. Event files were calibrated and cleaned by applying standard filtering criteria with the `xrtpipeline` task, using the latest calibration files available in the Swift CALDB distributed by HEASARC. For spectrum extraction, we considered events in the 0.3–10.0 keV energy range using the `xselect` tool. After visually inspecting the images to center the extraction region at the coordinates of the optical counterpart, we extracted the source signal within a circular region of radius 20 pixels (~47″), covering ≈90% of the XRT point-source function. For the background spectrum, an annular region centered on the source with an inner radius of 40 pixels (~1$'$.5) and an outer radius of 80 pixels (~3$'$.1) was used.

Following standard analysis procedures, we created both the exposure maps and the ancillary (ARF) files using the `xrtexpomap` and `xrtmkarf` tasks, respectively. For targets with multiple visits, such as T1 (a) and T1 (b), we then combined these files using the xselect and ximage tools to build a single event file and a single exposure map for each observation.

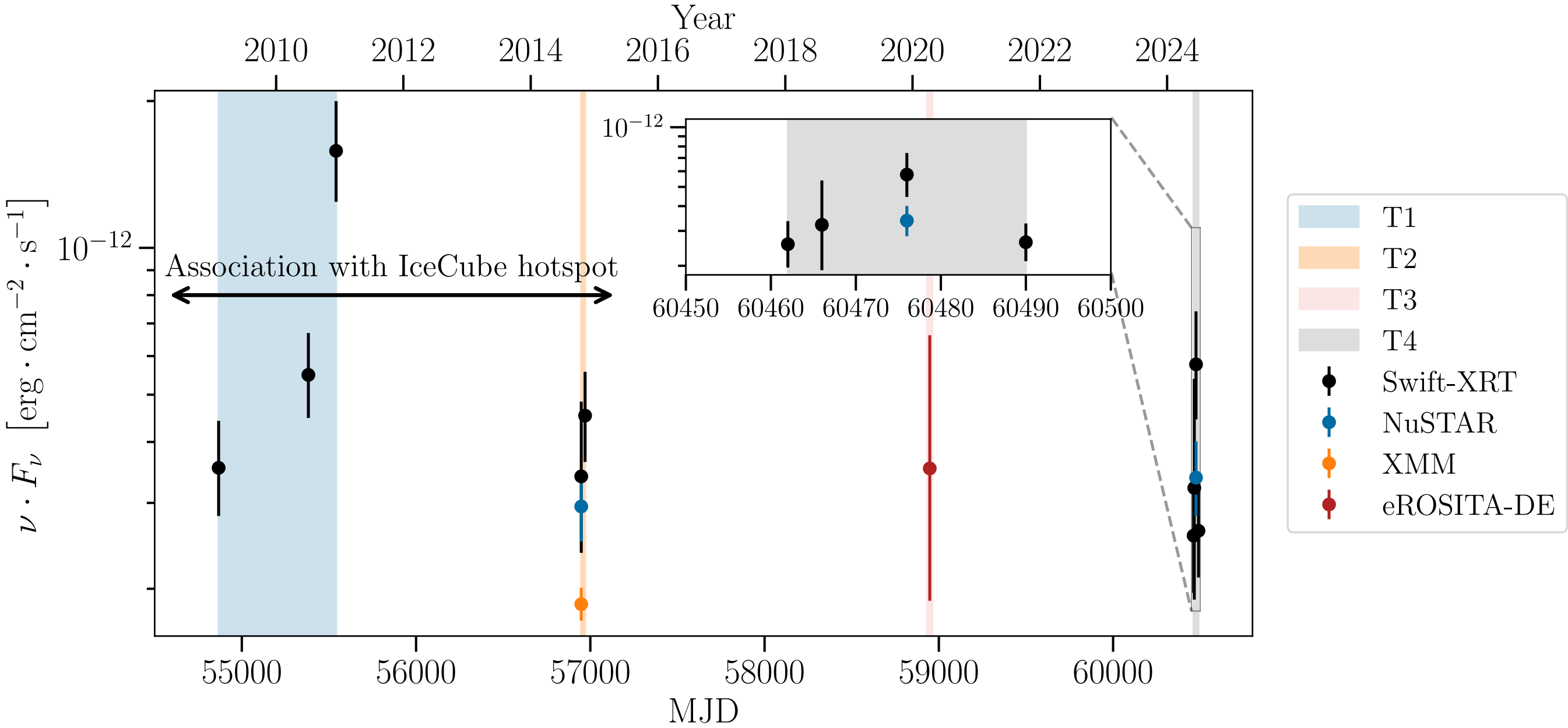

**Figure 1.** Light curve of 5BZB J0630-2406 with measurements from Swift-XRT, XMM-Newton, NuSTAR, and eROSITA for epochs reported in Table 1. Fluxes are derived in the 2.0–10.0 keV energy range. A double-headed arrow marks the association of 5BZB J0630-2406 with an IceCube neutrino hotspot over the period 2008–2015 (Paper I).

### 2.2. XMM-Newton and NuSTAR

For the XMM-Newton observation, the three EPIC instruments—two MOS cameras and one PN camera—were operated in full-frame mode (L. Strüder et al. 2001; M. J. L. Turner et al. 2001). All three EPIC instruments used the medium filter to prevent optical contamination from point sources as bright as $m_V = 6$–9. The observation data files (ODFs) were obtained from the XMM-Newton Science Archive and analyzed using the Standard Analysis System (SAS; v.21.0.0) following standard analysis threads. For this analysis, only the data from the PN camera were utilized due their higher signal-to-noise ratio. The epproc standard pipeline was used to properly process and correct the EPIC PN ODFs.

With the setup of the SAS environment, high-background events were filtered out, producing cleaned event files for spectral analysis. In particular, for this observation, a light curve was created to check for flaring high-background periods, including only single events with energy between 10 and 12 keV to avoid mistaking hot pixels, which show unusually high signals, for very high-energy events. Selecting a good time interval (GTI) was necessary to filter out periods of high background activity. For this analysis, the GTIs were chosen by considering only the periods where the observed rate was less than 0.4 counts $s^{-1}$.

For the spectral analysis obtained from XMM-Newton the source was positioned near the edge of a charge-coupled device (CCD). To avoid complications, such as including the CCD gap or parts of a neighboring CCD in the background extraction, we defined the background using a circular region on the same CCD. This region was placed at a similar distance from the readout node in a region free from any sources. This approach ensures a more accurate background extraction by minimizing potential contamination and aligning the background conditions with those of the source region. The source and background spectra were extracted from a 30″ circular region (≈90% of the encircled energy fraction at 1.5 keV) centered at the optical position of the source and from a nearby region (∼80″ separation) that was visually inspected to avoid contamination, respectively. Following this, the rmfgen and arfgen tasks were used to generate the redistribution matrix file and the ARF. For further spectral analysis, the spectrum was rebinned using the specgroup task to ensure at least one count for each background-subtracted spectral channel.

For the NuSTAR observations, held in SCIENCE mode, the data from both Focal Plane Modules (FPM) A and B were processed using the NuSTAR Data Analysis Software (NUSTARDAS) v.2.1.2 (F. A. Harrison et al. 2013). For this work, we used the data acquired by both cameras. The raw event files were calibrated by the nupipeline script, using the response file from the CALDB v.20240701. After a visual inspection of the event files, the source energy spectrum was extracted from a circular region centered at the optical position of the source with a radius of ∼50″, while the background spectrum was obtained from an annular region with inner and outer radii of ∼2.′0 and ∼3.′ 3 respectively. Using nuproducts scripts, we then generated source and background spectral files, along with the corresponding ARF and redistribution matrix files.

### 2.3. eROSITA-DE

The Spektrum-Roentgen-Gamma (SRG) observatory (R. Sunyaev et al. 2021), houses two principal instruments: eROSITA and ART-XC. eROSITA is an all-sky instrument that represents a major step forward in the analysis of the soft X-ray sky (i.e., at energies below 3–5 keV) with respect to previous facilities, such as ROSAT (P. Predehl et al. 2021). This instrument comprises seven X-ray telescope modules (TM1–TM7), each aligned in parallel, with a field of view of approximately 1° in diameter. These modules contain 54 nested mirror shells, and the performance of the system is defined by parameters such as effective area, vignetting function, and point-spread function. The last of these has achieved an average spatial resolution of about 30″ in survey mode, as initially analyzed in-flight (A. Merloni et al. 2024).

For the analysis of eROSITA data, we utilized the Data Release 1 (DR1) archive, which encompasses data from the first six months of the SRG/eROSITA all-sky survey (eRASS1). A comprehensive explanation of the data processing (pipeline version c001) is available in Section 3 of H. Brunner et al. (2022). Through the catalog search page,[10] we accessed individual products for the target, including source and background spectra, their respective ARFs and RMFs, a source light curve, and event lists for both source and background (iauname: 1eRASS J063136.5-240950). Specifically, these products for each active telescope module and combined TM configurations were generated using the eSASS task `srctool`, taking into account flaring events during the observation, e.g., `gti=FLAREGTI`.

In addition to the seven TMs on board eROSITA, a configuration is achieved by combining TMs with and without on-chip filters for the production of specific source products. For example, TM8 incorporates the five cameras equipped with on-chip filters (TM1, TM2, TM3, TM4, and TM6), while TM9 encompasses the remaining two cameras that lack this feature. Notably, the TM9 detectors experience time-variable light leaks, impacting their performance and calibration at the softest energies (refer to P. Predehl et al. 2021 for more details).

## 3. Spectral Analysis

For proper spectral modeling, binning was set to at least one count per spectral channel, enabling the use of C-statistics (W. Cash 1979). The binning was performed with `grppha` for Swift and NuSTAR data, while `specgroup` was used for XMM-Newton. This choice is beneficial as C-statistics, for low-count data, are derived from the likelihood function for Poisson-distributed data. The spectra are fitted in XSPEC (v12.14.0h), estimating the uncertainties at $1\sigma$ confidence (K. A. Arnaud 1996). In the fit, the Galactic absorption column density toward 5BZB J0630-2406 is fixed at $N_{\rm H,gal} = 7.5 \times 10^{20}\,{\rm cm}^{-2}$ (P. M. W. Kalberla et al. 2005). We fixed the metal abundance to solar metallicity using the abundances from J. Wilms et al. (2000), while the photoelectric cross sections for all absorption components are those derived by D. A. Verner et al. (1996). We also include an absorption component at the redshift of the source ($N_{\rm H,ISM}$, fixing $z = 1.239$) to account for contribution from the interstellar medium (ISM), as found in Paper TI.

As noted, only epoch T2 includes simultaneous observations from Swift, XMM-Newton, and NuSTAR, while T4 has joint observations with Swift and NuSTAR. For these two epochs, we applied a multiplicative constant in the model to account for cross-calibration uncertainties among different instruments (K. K. Madsen et al. 2015). Specifically for the analysis of data sets, the XMM-Newton PN spectrum is fitted in the 0.3–10 keV band, while the two NuSTAR FPMA and FPMB spectra are fitted in the 3.0–15.0 keV band. For this and the other epochs where Swift data are available, we considered data within the 0.3–10 keV energy range. Figure 1 shows the intrinsic X-ray fluxes at the different epochs.

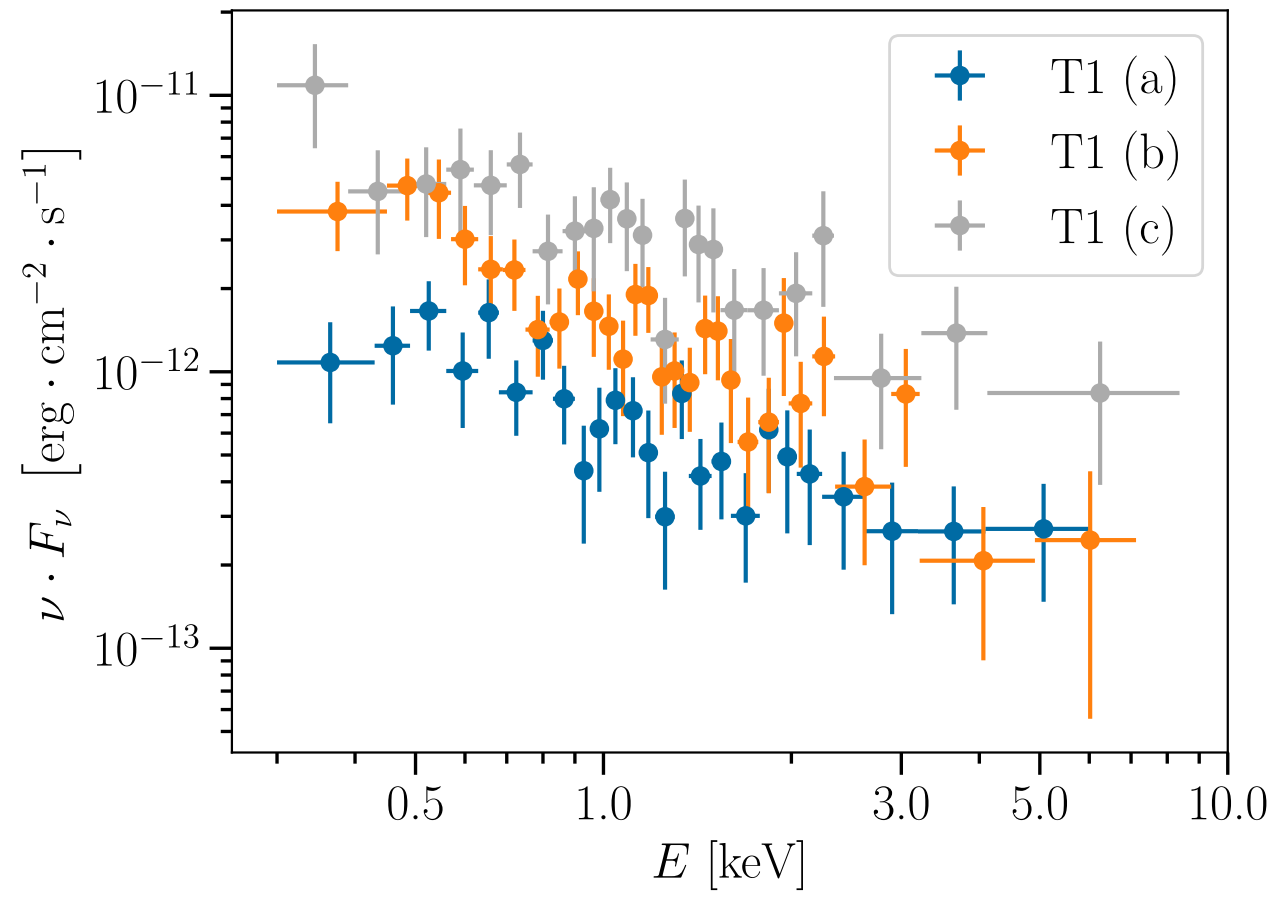

**Figure 2.** Absorption-corrected spectra for each of the T1 Swift-XRT visits analyzed. The best-fit models used to compute the fluxes are those reported in Table 2.

**Table 2**
Best-fit Parameters from the X-Ray Analysis of Swift-XRT Observations during T1

| Period | T1 (a) | T1 (b) | T1 (c) |
|---|---|---|---|
| Rate | 3.4 ± 0.3 | 6.2 ± 0.4 | 12.3 ± 1.0 |
| $N_{\rm H,ISM}$ | $4.1^{+1.8}_{-1.7}$ | $5.9^{+1.7}_{-1.6}$ | $4.2^{+2.0}_{-1.9}$ |
| Γ | $2.84^{+0.23}_{-0.22}$ | $3.13^{+0.22}_{-0.21}$ | $2.84^{+0.25}_{-0.24}$ |
| Norm | $4.5^{+0.7}_{-0.6}$ | $10.2^{+1.3}_{-1.2}$ | $20.0^{+3.3}_{-2.8}$ |
| $F_{2.0-10.0\ {\rm keV}}$ | $3.5^{+0.9}_{-0.7}$ | $5.5^{+1.2}_{-1.0}$ | $15.8^{+4.1}_{-3.4}$ |
| C-stat./dof | 97.4/122 | 105.8/139 | 80.1/111 |

**Note.** The units of $N_{\rm H,ISM}$ are $10^{21}\,{\rm cm}^{-2}$, the normalization is in units of $10^{-4}$ photons keV$^{-1}$ cm$^{-2}$, rates are in units of $10^{-2}$ counts s$^{-1}$, and the intrinsic flux in the 2.0–10.0 keV range is in units of $10^{-13}$ erg s$^{-1}$ cm$^{-2}$.

### 3.1. Epoch T1

Archival observations from Swift-XRT are available for T1 epochs. Using standard X-ray analysis procedures, we model the spectra with the simplest spectral shape, a power law, fixing the value of Galactic absorption, and considering an absorption component from the ISM (`phabs*zphabs*pow`). The spectra, as well as the result of the fit, can be seen in Figure 2 as well as in Table 2. The other models tested, including a broken power law or a log-parabola, provided no evidence for a further component that could improve the fit. This result is consistent with our expectations given the SED of the source. The observed X-rays trace the declining portion of the synchrotron hump in the SED, resulting in a simple power-law shape in the X-ray band, consistent with the expected decrease in synchrotron emission. The recovered photon indices are consistent within their uncertainties, further confirming the simple power-law model in describing the X-ray spectra during this first epoch.

### 3.2. Epoch T2

Two observations of 5BZB J0630-2406 were conducted during T2. The first, T2 (a), is a joint observation with Swift-XRT, XMM-Newton, and NuSTAR, which was analyzed in

[10] https://erosita.mpe.mpg.de/dr1/erodat/catalog/search/

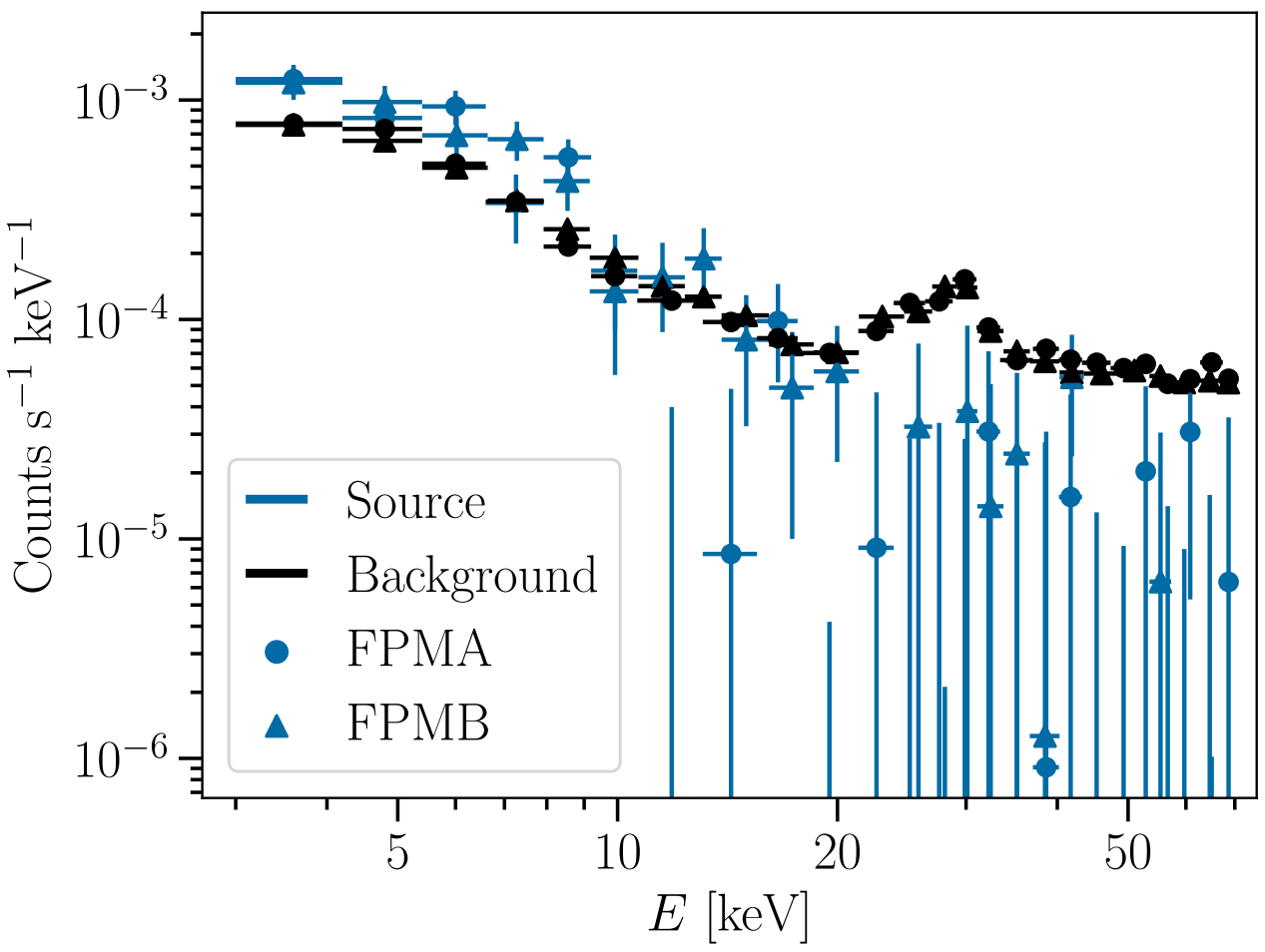

**Figure 3.** Rates for both cameras of NuSTAR during T2 (a). For both NuSTAR cameras, the background rate becomes dominant over the source rates above approximately 15.0 keV. The data have been rescaled to account for the differences in extraction areas between the source and background regions.

**Table 3**
Best-fit Parameters for T2 (a) and T2 (b) Using Both Power-law (PWL) and Broken Power-law (BKN) Models

| PWL Model | | |
|---|---|---|
| Parameter | T2 (a) | T2 (b) |
| $N_{\mathrm{H,ISM}}$ | $2.7^{+0.5}_{-0.5}$ | 5.5[a] |
| $C_{\mathrm{XMM\text{-}FPMA}}$ | $1.7^{+0.3}_{-0.2}$ | … |
| $C_{\mathrm{XMM\text{-}FPMB}}$ | $1.9^{+0.3}_{-0.3}$ | … |
| $C_{\mathrm{XMM\text{-}XRT}}$ | $0.8^{+0.1}_{-0.1}$ | … |
| $\Gamma$ | $3.12^{+0.08}_{-0.08}$ | $2.69^{+0.17}_{-0.16}$ |
| Norm | $4.2^{+0.2}_{-0.2}$ | $4.7^{+0.4}_{-0.4}$ |
| $F_{2.0-10.0\ \mathrm{keV}}$ | $2.3^{+0.2}_{-0.2}$ | $4.5^{+1.0}_{-0.9}$ |
| C-stat./dof | 901.4/980 | 99.1/105 |
| BKN Model | | |
| Parameter | T2 (a) | T2 (b) |
| $N_{\mathrm{H,ISM}}$ | $3.8^{+0.7}_{-0.6}$ | 5.5[a] |
| $C_{\mathrm{XMM\text{-}FPMA}}$ | $1.4^{+0.3}_{-0.2}$ | … |
| $C_{\mathrm{XMM\text{-}FPMB}}$ | $1.7^{+0.4}_{-0.3}$ | … |
| $C_{\mathrm{XMM\text{-}XRT}}$ | $0.8^{+0.1}_{-0.1}$ | … |
| $\Gamma_1$ | $3.36^{+0.13}_{-0.12}$ | $3.28^{+0.30}_{-0.27}$ |
| $E_{\mathrm{break}}$ | $2.6^{+1.1}_{-0.5}$ | $1.5^{+0.3}_{-0.3}$ |
| $\Gamma_2$ | $2.63^{+0.16}_{-0.16}$ | $1.82^{+0.38}_{-0.38}$ |
| Norm | $4.4^{+0.2}_{-0.2}$ | $4.2^{+0.5}_{-0.5}$ |
| $F_{2.0-10.0\ \mathrm{keV}}$ | $2.4^{+0.3}_{-0.3}$ | $7.9^{+2.6}_{-2.0}$ |
| C-stat./dof | 888.8/978 | 91.0/103 |

**Note.** The units of $N_{\mathrm{H,ISM}}$ are $10^{21}$ cm$^{-2}$, normalization is in units of $10^{-4}$ photons keV$^{-1}$ cm$^{-2}$, the intrinsic flux in the 2.0–10.0 keV band is in units of $10^{-13}$ erg s$^{-1}$ cm$^{-2}$, and $E_{\mathrm{break}}$ is in keV.
[a] Value that was frozen during the fitting process.

Paper TI. An additional observation by Swift-XRT was carried out approximately one month later, labeled T2 (b). These were taken toward the end of the IceCube observations that led to the association of 5BZB J0630-2406 with a neutrino hotspot. In this study, we revisit the analysis performed in Paper TI, expanding upon our earlier findings and including the additional XRT observations. Given the potential differences in selecting source and background regions, this leads to minor variations in parameter values while still supporting the overall conclusions of the original work.

### 3.2.1. *Epoch T2 (a): Joint Swift-XRT/XMM-Newton/NuSTAR Observations*

For epoch T2 (a), 5BZB J0630-2406 was observed simultaneously by XMM-Newton, NuSTAR, and Swift-XRT. For the XMM-Newton observation, the selection for the GTI resulted in a cleaned event file of 3.8 ks, i.e., $\sim$40% of the overall observation. As with the T1 observations, the data were fitted simultaneously accounting for both Galactic and ISM absorption. A cross-normalization constant was applied between the different instruments to account for differing observed rates due to the intrinsic characteristics of the telescopes (e.g., effective area). Both Swift and XMM-Newton points were fitted in the 0.3–10 keV band, while the NuSTAR ones were fitted between 3.0 and 15 keV. For this observation, the extracted spectrum was found to be background-dominated above 15.0 keV, as can be seen in Figure 3. Therefore, data above this energy were excluded from the analysis.

To describe the observed spectrum, we first tested a power law, and then we assumed a broken power-law, which is a model also commonly used to describe the X-ray spectra of blazars (A. Comastri et al. 1997). The results of the power-law and broken power-law fits, including the derived photon indices and break energy, are summarized in Table 3. Considering the difference between the fit statistics and the degrees of freedom between the two models, the broken power-law model is statistically preferred at a confidence level exceeding 99.8% and a significance level above 3.1$\sigma$. Despite minor differences in parameter values from Paper TI, likely due to the inclusion of XRT observations and small variations in selection of source and background regions, our results are consistent with those findings and support the same conclusion. The break energy is well constrained in this model, as shown in Figure 4, where the best-fit value for $E_{\mathrm{break}}$ appears as the global minimum. The lower panel of Figure 4 displays the contour plot for $E_{\mathrm{break}}$ and $N_{\mathrm{H,ISM}}$, highlighting the independence of these parameters. The modeled spectrum with the spectral break is shown in Figure 5.

This observation was also modeled excluding the Swift data, resulting in a fit equivalent to that obtained when considering all instruments. In other words, the XMM-Newton observations, with their higher signal-to-noise ratio than Swift, contribute more significantly to the total fit statistic. Additionally, we tested modeling the spectrum including only data from XMM-Newton and found minor evidence of an additional component in the spectrum (i.e., $\lesssim 1.4\sigma$). Based on this finding, observations in both the soft and hard X-ray energy bands are crucial to pinpoint additional components at a statistically significant level.

### 3.2.2. *Epoch T2 (b): Further Evidence for an Additional Component*

We analyzed the T2 (b) observation performed with XRT, which displays flux levels comparable to T2 (a) while benefiting from twice the exposure of that epoch. We proceeded by initially modeling the data with a power law, and fixing the ISM absorption component to the value previously found $N_{\mathrm{H,ISM}} = 5.5 \times 10^{21}$ cm$^{-2}$. The best-fit photon index is $\Gamma = 2.69^{+0.17}_{-0.16}$, indicating a considerable hardening of the spectra compared to the results found in T2 (a) for the power-law fit. Figure 5 shows the spectrum derived from this model. In this case, the best-fit statistics are C-stat = 99.1 with 105 degrees of freedom. At first glance, the excess residuals at low X-ray energies suggest considering a model that accounts for soft excess, which is typically done by adding a blackbody component. In AGN, this excess is often attributed to the declining part of a "big blue bump" (D. B. Sanders et al. 1989; J. N. Bregman 1990), observed rising in the optical–UV spectrum and indicative of accretion-disk emission (F. Tombesi et al. 2015). However, this is not the case for blazars as their emission is typically dominated by the jet. We therefore fit the data using a broken power-law model. This model yields a C-stat of 91.0 with 103 degrees of freedom, indicating a statistical preference over the simple power law at the $\sim 2.4\sigma$ level, consistent with our findings from T2 (a). The fit results for T2 (b) are shown in Table 3. This result complements the earlier findings, where the higher signal-to-noise ratio of the combined data allowed for stronger constraints in the spectral break. Furthermore, for the broken power-law model of both T2 observations, while the fitted value for $\Gamma_1$ is fully compatible, the fitted value for $\Gamma_2$ in T2 (b) is noticeably harder than that in T2 (a). This difference could reflect variations in the data quality or a more complex mechanism at play. These spectral properties will be further discussed in the following sections.

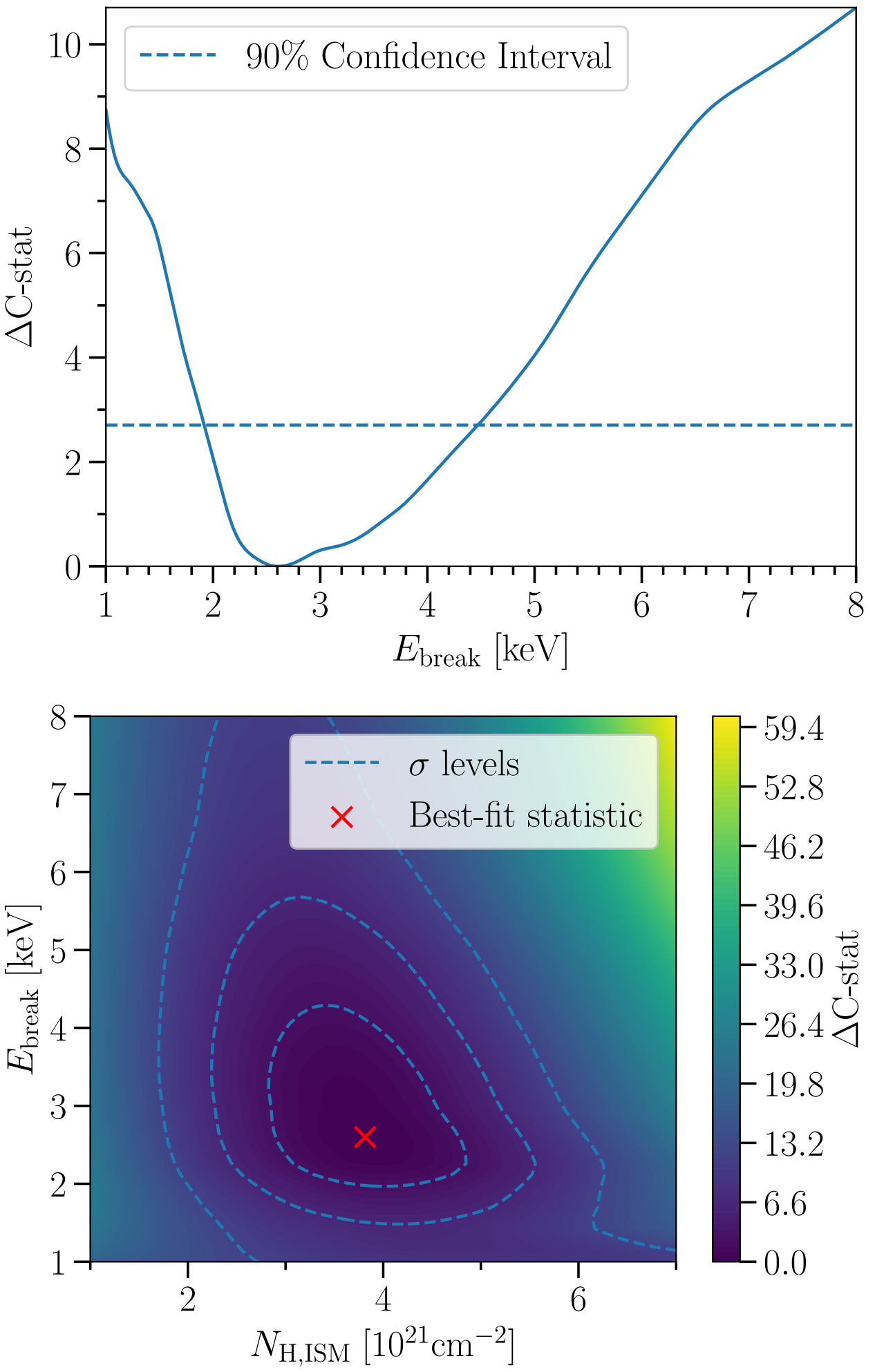

**Figure 4.** Likelihood profiles for the joint observation in T2 (a). Top: contour plot for $E_{\rm break}$ in the broken power-law model, with the blue dashed line representing the 90% confidence region. Bottom: profiles showing the ISM column density on the $x$-axis and $E_{\rm break}$ on the $y$-axis. The dotted lines mark the $1\sigma$, $2\sigma$, and $3\sigma$ levels, and the cross indicates the best-fit values.

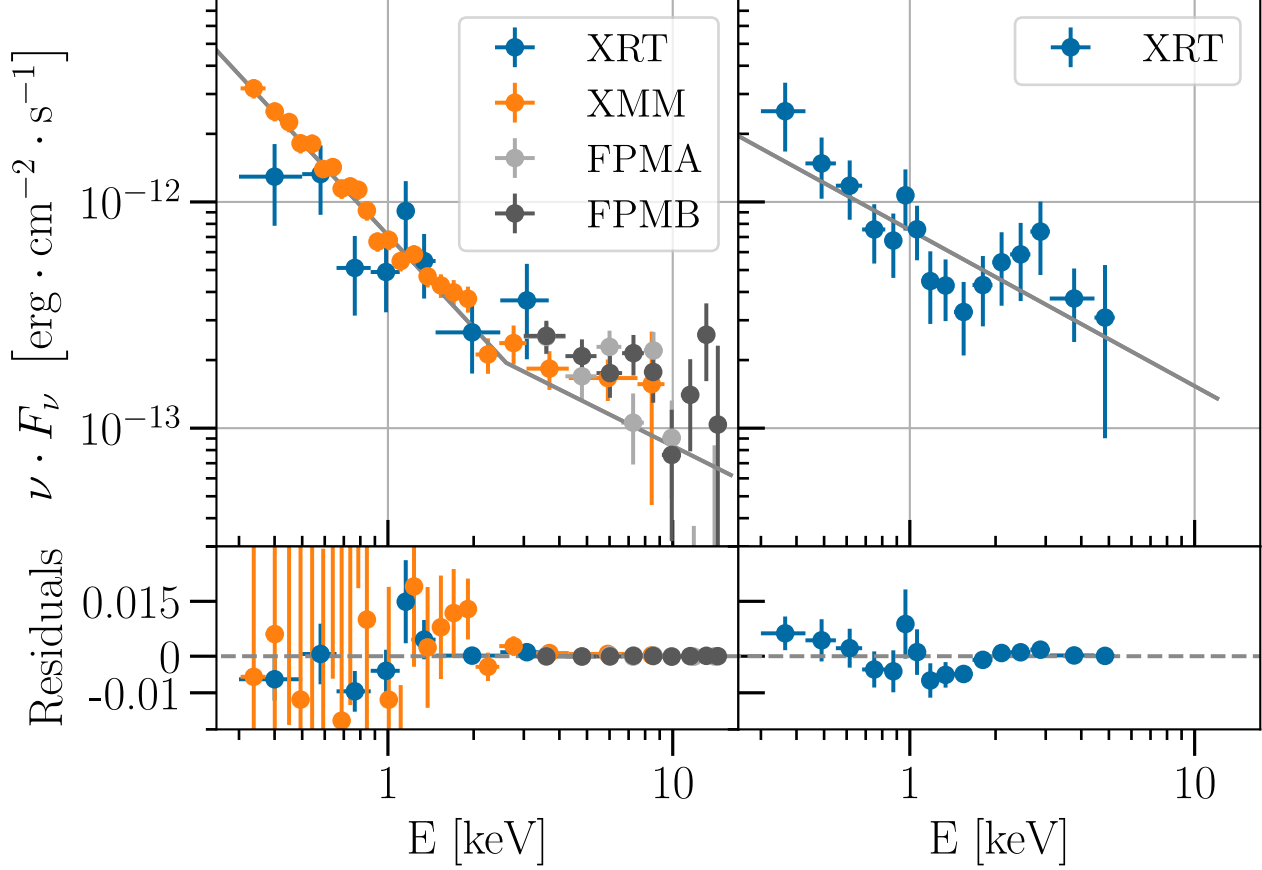

**Figure 5.** Spectra and residuals from both observations during T2. Left: T2 (a) joint observation is best fitted with a broken power-law model. Right: T2 (b) spectrum assuming a simple power-law model. The residuals have units of counts s$^{-1}$ keV$^{-1}$.

### 3.3. Epoch T3

As described in Section 2.3, we utilized the data products from the eROSITA-DE data archive for DR1 to perform a spectral analysis of the source approximately seven years after its observation during the T2 epoch. For this analysis, we used the calibrated source products from TM8 and TM9, covering the 0.2–10.0 keV energy range. This broad energy coverage, combined with the high sensitivity of eROSITA, particularly below 2.0 keV, allowed us to better constrain the ISM absorption component by leaving this parameter free to float during the fitting process. Due to the light leaks mentioned before, similar to A. Veronica et al. (2024), distinct lower energy thresholds were employed for the modeling of the data. Specifically, for TM8 and TM9 we used lower limits of 0.2 keV and 1.0 keV, respectively. We modeled the data up to the 2.3 keV band, as this is the range of maximum effective area before this sharply declines at energies $\gtrsim$3 keV (S. Marchesi et al. 2025).

By independently fitting the TM8 and TM9 spectra with an absorbed power-law model, we found a best-fit photon index of $\Gamma = 2.91^{+0.60}_{-0.55}$ and an ISM absorption column density of $N_{\rm H,\,ISM} = 3.1^{+3.4}_{-2.9} \times 10^{21}\,{\rm cm^{-2}}$. By extrapolating the model to higher energies, we measured the flux to be $F_{2.0-10.0\,{\rm keV}} = 3.5^{+3.1}_{-1.6} \times 10^{-13}\,{\rm erg\,s^{-1}\,cm^{-2}}$. This model resulted in a fit with a fit statistic value of C-stat = 63.8 with 79 degrees of freedom. Due to the low statistics, no further models were tested.

### 3.4. Epoch T4: Dedicated Observations

Based on the compelling findings from previous epochs and Paper TI, we performed follow-up observations with Swift and NuSTAR, as part of the Swift cycle-20 program. A joint Swift-XRT and NuSTAR observation was performed in 2024. This was accompanied by three additional Swift-XRT visits, about two months apart from each other, to access medium-term variability in the target. Similar to the NuSTAR data obtained during T2 (a), the extracted spectrum for our source was found to be background-dominated above 15 keV. The new NuSTAR

**Table 4**
Best-fit Parameters from Analysis of the X-Ray Observations during T4

| Instrument | Swift | Swift | Swift–NuSTAR | Swift |
|---|---|---|---|---|
| Epoch | T4 (a) | T4 (b) | T4 (c) | T4 (d) |
| Rate (XRT) | $3.5 \pm 0.4$ | $3.1 \pm 0.4$ | $4.0 \pm 0.3$ | $3.2 \pm 0.3$ |
| Rate (FPMA+FPMB) | | | $0.8 \pm 0.1$ | |
| $C_{\text{XRT-FPMA}}$ | … | … | $0.5^{+0.3}_{-0.2}$ | … |
| $C_{\text{XRT-FPMB}}$ | … | … | $0.6^{+0.4}_{-0.2}$ | … |
| $N_{\text{H,ISM}}$ | 5.5[a] | $4.1^{+3.9}_{-3.3}$ | $3.1^{+1.8}_{-1.6}$ | 5.5[a] |
| $\Gamma$ | $3.25^{+0.21}_{-0.20}$ | $3.18^{+0.56}_{-0.52}$ | $2.71^{+0.20}_{-0.19}$ | $3.17^{+0.17}_{-0.17}$ |
| Norm | $5.5^{+0.5}_{-0.5}$ | $6.4^{+2.0}_{-1.5}$ | $7.4^{+1.1}_{-1.0}$ | $5.1^{+0.4}_{-0.4}$ |
| $F_{2.0-10.0\ \text{keV}}$ | $2.6^{+0.8}_{-0.6}$ | $3.2^{+2.2}_{-1.3}$ | $7.0^{+0.6}_{-0.6}$ | $2.6^{+0.6}_{-0.5}$ |
| C-stat./dof | 88.1/81 | 37.7/43 | 272.1/285 | 105.4/111 |

**Note.** The units of $N_{\text{H,ISM}}$ are $10^{21}$ cm$^{-2}$, the normalization is in units of $10^{-4}$ photons keV$^{-1}$ cm$^{-2}$, rates are in units of $10^{-2}$ counts per second, and the intrinsic flux in the 2.0–10.0 keV range is in units of $10^{-13}$ erg s$^{-1}$ cm$^{-2}$.
[a] Value that was frozen during the fitting process.

observation is ∼3.5 times shorter than during T2 (a). We address this limitation in Section 4.2.

The best-fit parameters for an absorbed power-law model are listed in Table 4, and the corresponding spectra are shown in Figure 6. The fit statistic values are close to the degrees of freedom, indicating that the model describes the data well within the limits of statistical uncertainties. The epoch with joint Swift-XRT/NuSTAR observations allowed us to constrain $N_{\text{H,ISM}}$, leading to consistent results with previous measurements. During this new observation at T4, we observed a hardening of the spectrum, suggesting a potential change in the spectral shape. We tested for the presence of a spectral break assuming a broken power-law model. No significant improvement in the model is found when modeling a broken power law or any other multicomponent models used, e.g., log-parabola, even though the source was in a similar flux state to that in T2, where this behavior was first observed. The lack of evidence for a spectral break, especially during T4 (c) where NuSTAR data are available to constrain the hard X-ray band, will be addressed in the next section.

## 4. Further Consistency Tests on the Additional X-Ray Component

### 4.1. Splitting the T2 (a) Data

To further test the presence of an additional component in the X-ray spectrum as observed during T2 (a), we analyzed subsets of the joint XMM-Newton and NuSTAR data, dividing the observations into two groups containing approximately half of the data from each mission. Spectral analyses of these subsets confirmed the presence of a break, with significance levels of $\sim 2\sigma$ and $\sim 3\sigma$ in the two groups, respectively. Furthermore, the best-fit parameters obtained from the subsets were fully consistent with those derived from the full T2 (a) data set. The persistence of the break across both subsets supports its genuineness.

### 4.2. Impact of Limited Statistics in T4

Despite the target being in a comparable flux state during T4 and T2 epochs, and the joint Swift–NuSTAR observation at T4 (c) providing hard X-ray coverage, no clear additional component in the X-ray emission is detected in T4. We performed a joint analysis of the observations during T2 (a) and T4. The results demonstrate that the break remains statistically significant at $>3.2\sigma$. Notably, the T2 (a) observations contribute approximately ∼70% of the total statistical weight, indicating that the lack of a clear break in T4 is likely due to insufficient data quality rather than an intrinsic absence of the feature. This finding reinforces the idea that the observed spectral break is a robust characteristic of the source, while highlighting the critical role of sufficient high-quality data for detecting such features.

### 4.3. Simulations with Exposure as of T4 Epoch

We use the XSPEC `fakeit` command to simulate spectra and assess the likelihood that an additional component is also present during the T4 epoch but remains undetected due to insufficient statistics. We produced $10^4$ simulated spectra for both the Swift and NuSTAR data sets, utilizing the response files from the actual T4 observations. For the XRT data, as shown in Figure 1, no significant variability was detected within the Swift-XRT energy range during the T4 timeframe, therefore we combined all XRT observations taken in 2024 to create a single spectrum with a total XRT exposure time of 13.5 ks. We ensured that all simulated XRT spectra had consistent exposure times for both the background and the signal to replicate the combined observation. For the NuSTAR data, we maintained the original exposure times from the actual observations. To generate the simulated spectra, we used as reference the best-fit broken power-law model from T2 (a), $\Gamma_1 = 3.36$, $\Gamma_2 = 2.63$, and $E_{\text{break}} = 2.6$ keV, and renormalized the flux to match the observed flux values during T4. We fitted both a power-law model and a broken power-law model to each simulated spectrum to infer the ΔC-stat for each case. To account for the different responses between the XRT and FPM cameras, we included a normalization constant in the model, as we did in the actual observations. The parameter space for the photon index ranged from [1.5, 5.0] for the soft energies and from [−2.0, 5.0] for the hard energy range with a break energy between [0.01, 20.0] keV. The remaining model parameters were allowed to vary freely during the fitting process. Throughout this process, we also kept track of whether the fits not only improved the C-statistic but also reliably recovered the input parameters, ensuring the results were not skewed by parameter degeneracies.

After running the simulations, we excluded trials that showed parameter values truncated at the boundaries, leaving us with a total of $9.2 \times 10^3$ valid simulations. The median values of the best-fit parameter distributions recovered are as follows: $\Gamma_{\text{PWL}} = 3.08 \pm 0.16$, indicating the photon index of the power-law model; $\Gamma_1 = 3.43^{+0.34}_{-0.29}$, representing the photon index before the spectral break; $\Gamma_2 = 2.35^{+0.49}_{-0.29}$, representing the photon index after the spectral break; and $E_{\text{break}} = 3.3 \pm 1.8$ keV, which denotes the energy at which the spectral break occurs. These values are consistent with the best-fit parameters used in the simulated spectra as reference from the T2 epoch, properly recovering the expected spectral characteristics.

Our objective is to determine the likelihood of recovering the spectral break at a confidence level equal to or higher than $3\sigma$, by fitting the data with a broken power-law model. To assess the significance of the break, we measured the

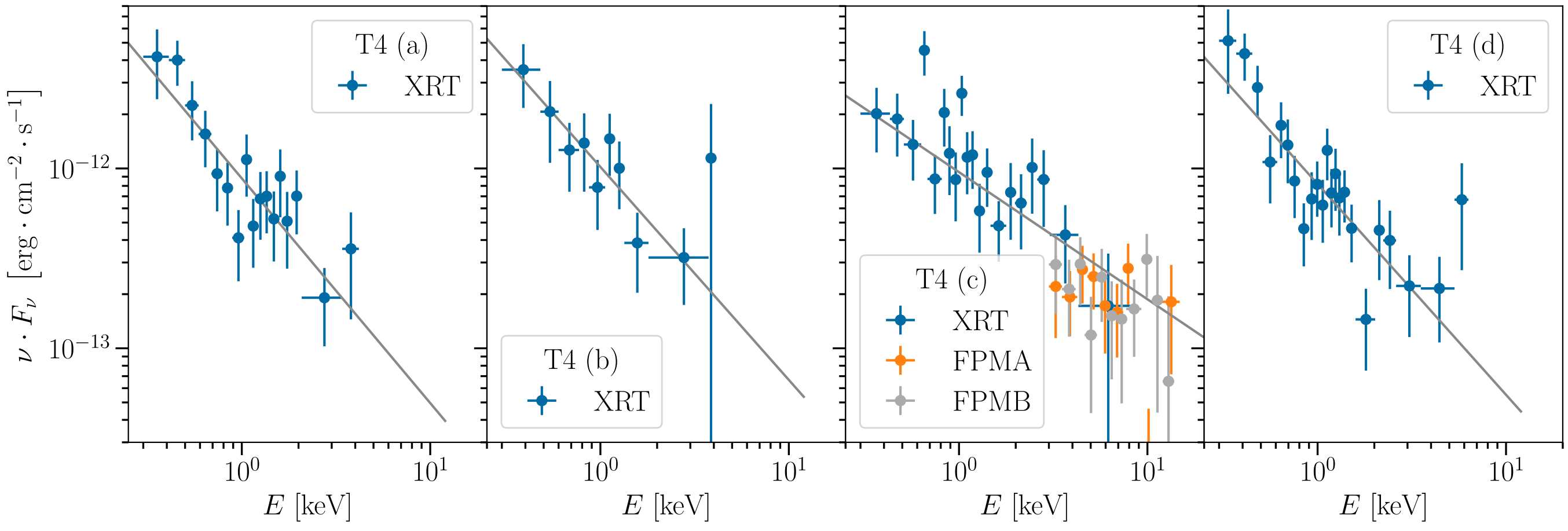

**Figure 6.** Modeled spectra from the observations taken during T4 with best-fit parameter values shown in Table 4.

ΔC-statistic between the fit statistics of the null hypothesis (simple power law) and the alternative hypothesis (broken power law). The distribution of the ΔC-stat values from these simulations is shown in Figure 7. The observed ΔC-stat between the models applied to the real data is 1.1, represented by a continuous pink line. In contrast, a dashed red line marks the $3\sigma$ improvement threshold for a one-tailed distribution with two degrees of freedom. The simulations indicate that the simulated spectra exceed the observed value of the ΔC-stat in 96% of the cases, while the ΔC-stat surpasses the $3\sigma$ threshold only 11% of the time. This suggests that, with an exposure equivalent to that of T4, while a spectral break may be present, it remains undetectable due to the limited signal-to-noise ratio in the T4 observational data. Therefore, we conclude that the presence of an additional component in that epoch cannot be ruled out.

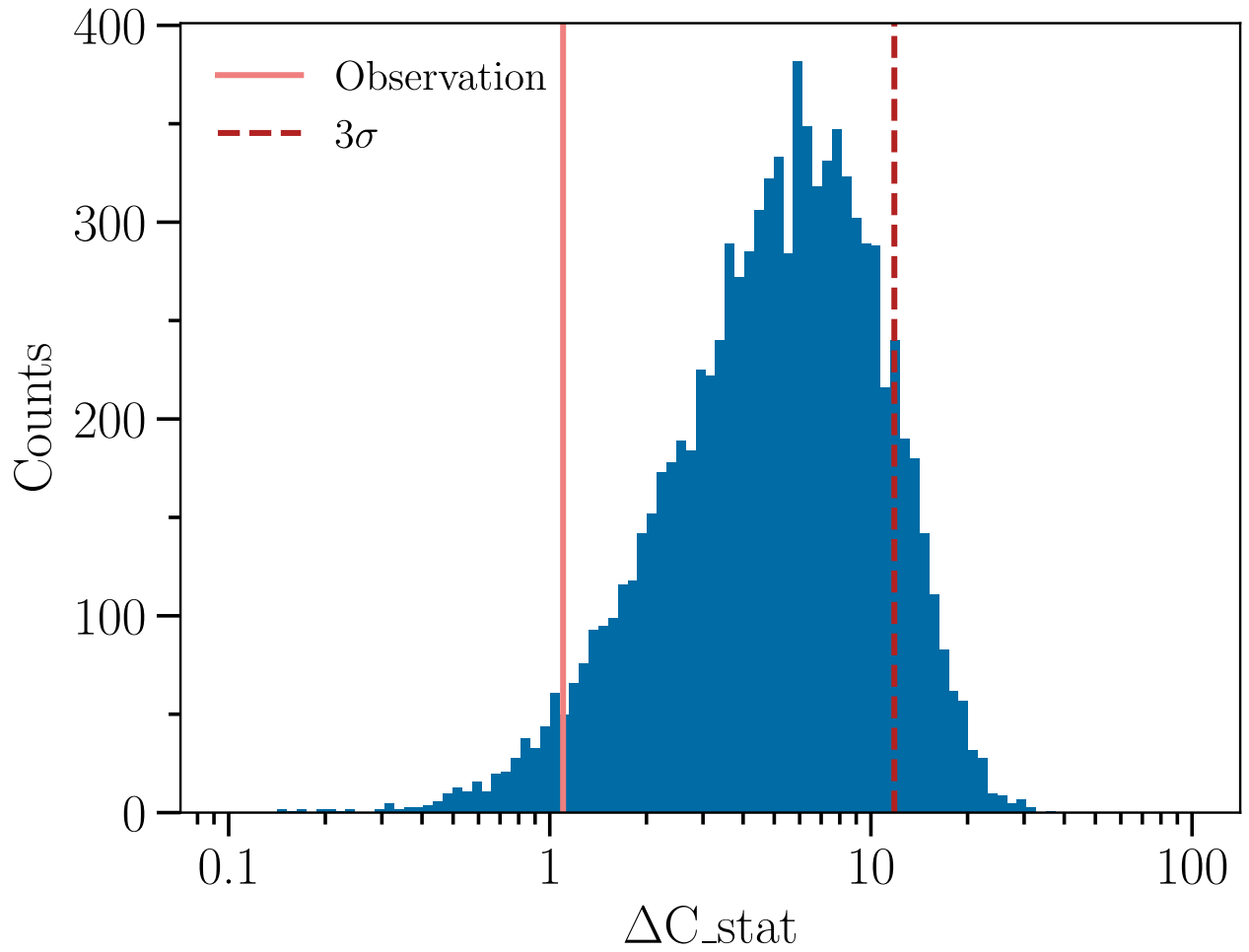

**Figure 7.** Distribution of ΔC-stat between a power law and a broken power law on $9.2 \times 10^3$ simulations. The solid pink line and the dashed red line represent the ΔC-stat recovered from the observation and the required value for a $3\sigma$ significance improvement of an additional component over the null hypothesis, respectively. From the distribution, we recover a greater significance of better modeling the data with a broken power law >95% of the time.

### *4.4. Exposure Required to Confirm or Rule out the Additional Component*

Finally, we estimated the exposure needed to confirm or rule out at high confidence the presence of an additional component with new observations. For this analysis, we renormalized the flux levels to match the most recent observations of the source while retaining the original response files and background data of T2 (a). Similarly to the previous test, the simulated spectra were generated using the best-fit broken power-law model from the T2 (a) epoch, and we measured the ΔC-statistic between the fit statistics of the null hypothesis (simple power law) and the alternative hypothesis (broken power law). We report the results of simulated, joint observations with XMM-Newton 33 ks and NuSTAR 80 ks exposures in Figure 8. The bottom panel shows the 68%–95%–99% confidence level contours for the two photon indices versus the break energy, highlighting that we would reliably measure any change in slope, if present. The simulations, after excluding cases where parameter values were truncated at boundaries, yielded a median ΔC-stat ∼ 32.2 (Figure 8 top). This corresponds to a detection significance of $\sim 5\sigma$, allowing us to confidently reject the null hypothesis of a simple power-law spectrum. More than 98% of the simulated spectra yield a ΔC-stat value exceeding the $3\sigma$ threshold, while over 60% surpass the $5\sigma$ level. This suggests that such combined observations will have a high likelihood of detecting or ruling out an additional component in the majority of cases.

## 5. Flux and Spectral Evolution

The multiepoch observations highlighted variations both in flux states and spectral properties of 5BZB J0630-2406. During the T1 epoch, the light curve shows an increasing X-ray flux, leading to a flare at T1 (c). In this preflaring and flaring state, the spectrum is well described by a simple power-law model. In contrast, the postflaring T2 epochs exhibit a spectral change. In T2 (a), simultaneous XMM-Newton, NuSTAR, and Swift-XRT observations reveal an additional spectral component beyond a power law, a feature that persists in T2 (b), where the Swift-XRT observation one month later shows the source in a similar low-flux state. At T3, eROSITA data have limited statistics to test models beyond a simple power law. Similarly, at T4, the source remains in a low state, consistent with a power-law shape. However, simulations suggest that, while the presence of an additional component cannot be confirmed, with the given exposure the limited signal-to-noise ratio prevents it from also being firmly ruled out. In the following sections we explore alternative explanations for the spectral behavior observed during T2 (b), and put these into the context of previous findings.

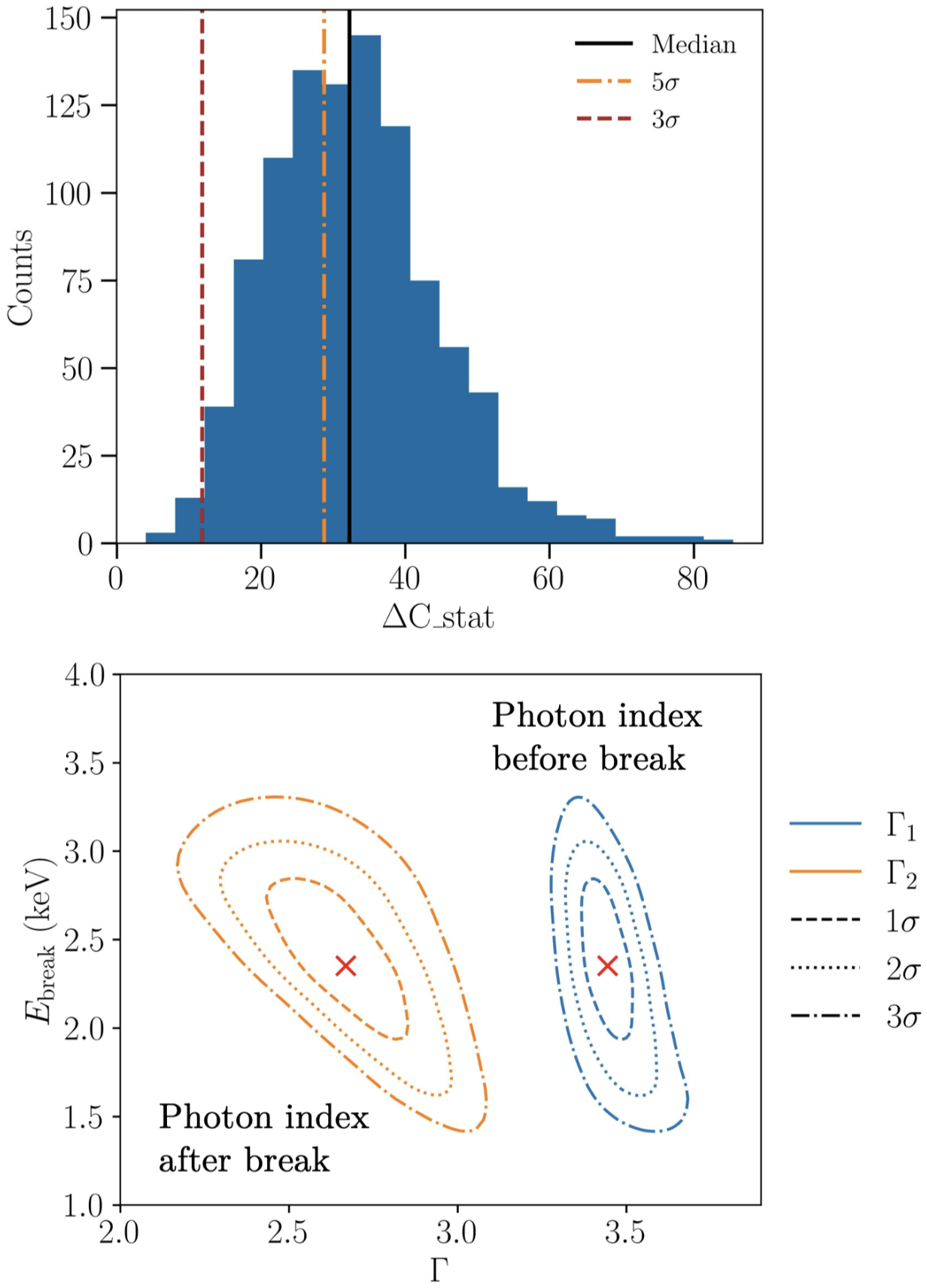

**Figure 8.** Results from the simulations needed to confirm or rule out the additional component. Top: distribution of $\Delta$C-stat for the $10^3$ simulated spectra. The $3\sigma$ and $5\sigma$ significance thresholds are marked as red dashed and yellow dashed-dotted lines respectively. From the distribution, we expect to recover a $3\sigma$ break significance >98% of the time, and a $5\sigma$ one >50% of the time. Bottom: the 68%, 95%, and 99% confidence contours of the two photon indices of the broken power-law model vs. the break energy.

## 6. Physical Origin of the Additional X-Ray Component

The evolving X-ray spectral properties of the blazar 5BZB J0630-2406 share similarities with the "changing-look" behavior typically observed in Seyfert galaxies. Additionally, various studies have reported diverse measurements characterizing the interaction between the jet and the nuclear environment in jetted AGN (see, e.g., P. Grandi & G. G. C. Palumbo 2004; X.-N. Sun et al. 2014; I. M. Mutie et al. 2025; S. Swain et al. 2025; and references therein). In the following, the working hypothesis is that while high states are primarily dominated by X-ray emissions from the powerful jet, lower jet states may allow us to observe additional contributions, such as emissions from a hot corona near the SMBH or effects of obscuration. In this scenario, we may observe various effects in X-rays, such as direct power-law emissions from the corona itself or neutral and ionized reflections of the coronal emissions. In this context, we use the term "corona" in a broad sense to describe a compact, central X-ray-emitting region, analogous to the coronae in Seyfert galaxies, but without assuming a specific physical model or detailed properties such as temperature or geometry. Another possibility is that the emission of the jet interacts with intervening material, and the latter causes the absorption of emissions from the jet. To investigate these possibilities, we test a set of models applied to the T2 (b) data set and then check the consistency of the findings with the T2 (a) epoch.

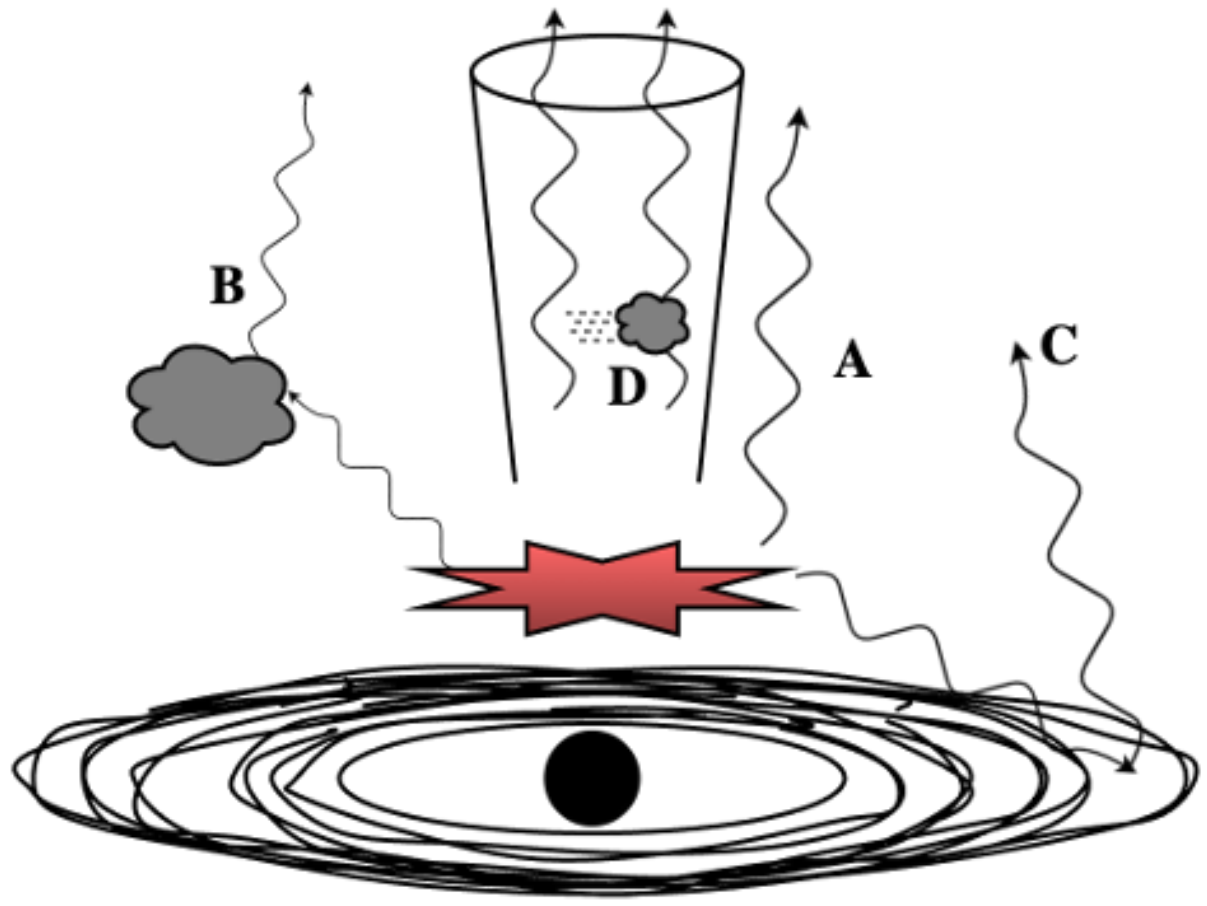

**Figure 9.** Sketch of the AGN illustrating the different scenarios tested to explain the observed X-ray spectrum in T2 (b). The central black hole is surrounded by an accretion disk with a corona (red star) located directly above it with optically thick clouds colored in gray. The figure includes the various geometries described in Section 6, as well as the direct emission from the jet, represented as a cone perpendicular to the disk.

We evaluated the following models: (A) a two-component power-law model representing emissions from the jet and corona; (B) `pexmon`, which parameterizes neutral reflection of the corona's emission (K. Nandra et al. 2007); (C) `xillver`, modeling ionized reflection in the accretion disk illuminated by the corona (J. García & T. R. Kallman 2010; J. García et al. 2011, 2013); and (D) `zxipcf`, which mimics partial covering of ionized absorption within the jet (L. Miller et al. 2006). For all of them, we considered the Galactic and the ISM absorption components by fixing their values to the same ones used for the simple and broken power-law models, i.e., $N_{\mathrm{H,gal}} = 7.5 \times 10^{20}\ \mathrm{cm}^{-2}$ and $N_{\mathrm{H,ISM}} = 5.5 \times 10^{21}\ \mathrm{cm}^{-2}$, as described in Section 3.2. A visual representation of the tested geometries is provided in Figure 9.

The main findings are summarized in Table 5, and the spectral models are presented in Figure 10. Next, we discuss the analysis and findings of each model, and we interpret them in the multimessenger context in Section 7.

### *6.1. Model A: Primary Coronal Continuum Emission*

We modeled the data of T2 (b) with a two-component power law, which allows us to mimic the emissions from the jet and the corona. The derived best-fit photon index for the corona, $\Gamma_{\mathrm{corona}} = 1.8^{+0.6}_{-0.9}$, is in line with typical values reported by M. Laurenti et al. (2022) in their X-ray spectral analysis of AGN with high Eddington ratio , $\Gamma_{\mathrm{corona}}$ spanning a range from 1.3 to 2.5. The values from this model are also consistent with those found in the broken power-law model for this same epoch T2 (b), reported in Table 3.

A two-component power-law model adequately described also the spectrum of the T2 (a) epoch. The recovered photon index in the soft X-ray band for the jet, $\Gamma_{\mathrm{jet}} \approx 3.3$, is in line with what is recovered from the analysis of T2 (b) with a turnover of the spectrum between 2 and 5 keV, further supporting the idea that we could be observing two separate phenomena: the jet and the corona.

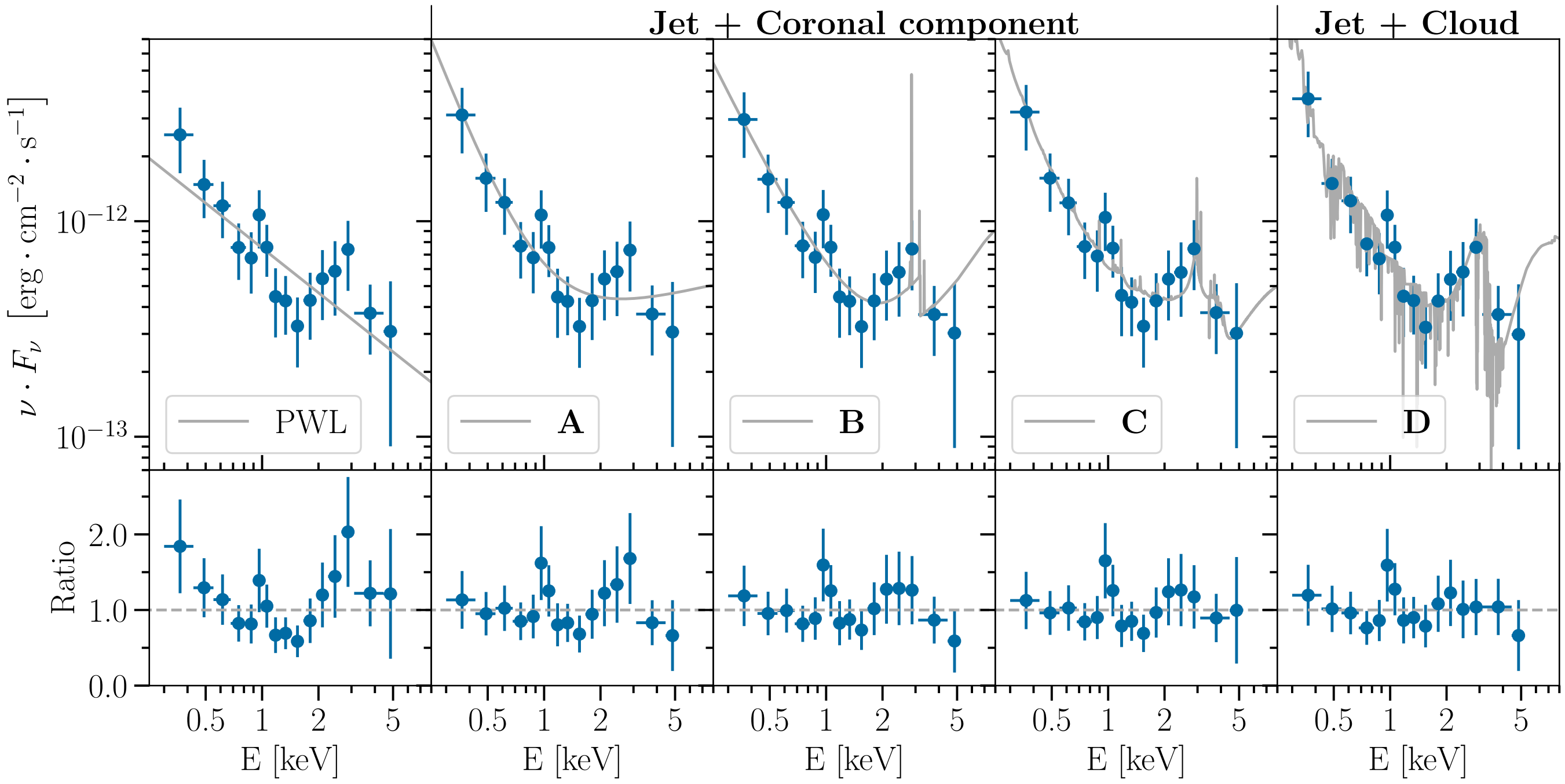

**Figure 10.** Spectra and ratio between the data obtained from Swift-XRT and the tested models in Table 5 for the Swift T2 (b) observation, plotted in the frame of the observer.

**Table 5**
Comparison of Parameters across Different Models Using the Month-apart Swift Observations at T2 (b)

| Model Parameter | Units | A<br>Coronal Continuum<br>PWL$_{jet}$+PWL$_{corona}$ | B<br>Neutral Reflection<br>PWL$_{jet}$+pexmon | C<br>Ionized Reflection<br>PWL$_{jet}$+xillver | D<br>Ionized Absorption<br>zxipcf*PWL$_{jet}$ |
|---|---|---|---|---|---|
| $N_{H,cloud}$ | $10^{23}$ cm$^{-2}$ | ⋯ | ⋯ | ⋯ | $16.8^{+1.1}_{-1.8}$ |
| $\Gamma_{jet}$ | ⋯ | $4.22^{+0.84}_{-1.82}$ | $3.84^{+0.56}_{-0.49}$ | $4.62^{+2.60}_{-1.14}$ | $3.78^{+0.32}_{-0.32}$ |
| Norm$_{jet}$ | $10^{-4}$ photons keV$^{-1}$ cm$^{-2}$ | $1.9^{+2.6}_{-1.9}$ | $2.4^{+1.0}_{-0.9}$ | $1.2^{+1.8}_{-1.1}$ | $114.5^{+84.1}_{-54.5}$ |
| $\Gamma_{corona}$ | ⋯ | $1.79^{+0.55}_{-0.86}$ | 2.5[a] | $1.94^{+0.34}_{-0.49}$ | ⋯ |
| Norm$_{corona/xillver}$ | $10^{-4}$ photons keV$^{-1}$ cm$^{-2}$ | $2.0^{+1.7}_{-1.5}$ | $1.4^{+0.4}_{-0.6}$ | $9.3^{+11.5}_{-2.8} \times 10^{-2}$ | ⋯ |
| $E_{cut}$ | keV | ⋯ | 200[b] | 200[b] | ⋯ |
| Incl. | deg | ⋯ | 5[b] | 5[b] | ⋯ |
| $z_{cloud/disk}$ | ⋯ | ⋯ | 1.3[b] | 1.3[b] | 1.3[b] |
| $\log X_i$ | ⋯ | ⋯ | ⋯ | $3.3^{+0.3}_{-0.4}$ | $2.8^{+0.1}_{-0.1}$ |
| $R$ | ⋯ | ⋯ | >20 | −1[b] | ⋯ |
| C-stat./dof | ⋯ | 92.2/103 | 90.4/103 | 88.9/102 | 88.7/103 |
| ΔC-stat./Δdof | ⋯ | 6.9/2 | 8.7/2 | 10.2/3 | 10.4/2 |
| Statistical significance | ⋯ | 2.1$\sigma$ | 2.5$\sigma$ | 2.4$\sigma$ | 2.8$\sigma$ |

**Notes.** Through all of the models, the Galactic and the ISM absorption components were fixed to the same values used for the simple and broken power-law models. In particular, for models B and C, the abundances of elements heavier than He, including the iron abundance, were set to solar values. Also, for model D, we set a completely covered source, i.e., with a covered fraction of 100%. In the table, $X_i$ is the ionization parameter of the reflector component (C. B. Tarter et al. 1969). In particular, the normalization for xillver follows the definition of T. Dauser et al. (2016, Equation (A.1)). The differences in the fit statistics and the degrees of freedom with respect to a single power-law model are shown as well as the significance in the improvement at the bottom of the table.

[a] Value that reached hard limits.

[b] Value that was fixed during modeling.

### 6.2. Model B: Neutral Reflection of Emissions of the Corona

A reflection component from the hot corona off optically thick material, such as a molecular torus, is typically observed in obscured AGN and Seyfert 2 galaxies. For example, the candidate neutrino emitter NGC 1068 shows strong reflection-dominated X-ray spectra. In our analysis, the neutral reflection component is modeled using pexmon, assuming an accretion disk with an inclination of 5°, based on expectations for a blazar. We set the redshift of the reflecting cloud to match that of 5BZB J0630-2406, and fixed the metal abundances, including iron, to solar values. The intrinsic emission from the corona is represented by a cutoff power law with an energy cutoff at $E_{cut} = 200$ keV, a typical value for AGN (M. Baloković et al. 2020).

The results of this model B, as reported in Table 5 and in Figure 10, show a power-law photon index for the jet of approximately $3.8^{+0.6}_{-0.5}$, consistent with the outcome of fitting a simple power law. This result is expected, as pexmon primarily models the hard portion of the spectrum, causing a softening effect on the primary power-law component. In the fit, the photon index of the pexmon component reaches its upper limit, $\Gamma_{\rm corona} = 2.5$, due to the reliance of the model on previously generated tables limited to $1.1 < \Gamma < 2.5$. Additionally, the reflection strength parameter, $R$, was found to be notably large ($R > 20$), resulting in the reflection dominating the emission above 2 keV. However, such a high value of $R$ suggests that this model may not provide a physically reliable description of the data, as it implies an unrealistically dominant reflection component. While studies of Seyfert 2 galaxies, such as those by S. Marchesi et al. (2022) and A. Pizzetti et al. (2022), effectively constrain the reflection parameter occurring on nuclear scales at the core of the AGN, this scenario is less plausible for a blazar like 5BZB J0630-2406.

For the T2 (a) epoch, results closely resemble those obtained for the T2 (b) data. The spectral fit yields a reflection strength parameter in the range $R \sim 0.2$–20 and a coronal photon index of $\Gamma_{\rm corona} = 2.5$. Additionally, the recovered jet spectral index of $\Gamma_{\rm jet} \sim 3.5$ is consistent with a two-component scenario for the observed X-ray SED.

This model provides a statistically significant improvement over a simple power-law fit, with a preference at the $>3\sigma$ level. While this scenario remains intriguing in the context of high-energy neutrino production, the inferred reflection strength and coronal properties remain somewhat difficult to reconcile with the expected physical conditions of the source.

### 6.3. Model C: Nonrelativistic Ionized Reflection

We tested ionized reflection in the accretion disk illuminated by the corona using the xillver model from the relxill package (T. Dauser et al. 2016). The xillver model simulates nonrelativistic reflection from an ionized accretion disk. The model we employed consists of a power-law component to account for the primary radiation from the jet, combined with reflected emission from a corona situated directly above the disk. This setup typically represents reflection from regions further from the central compact object, where general relativistic effects are minimal.

Given the characteristics of the source, we fixed the disk inclination at 5°, and set the cutoff energy of the coronal emissions to $E_{\rm cut} = 200$ keV with the metal abundances fixed to solar values, similarly to model B. We assume that there are no contributions from the direct component irradiating the disk by setting the reflective fraction $R = -1$. According to the best-fit model, the soft X-ray band is dominated by emission from the jet, while the excess above 2 keV is attributed to reflection from the corona, which exhibits a harder photon index than the jet.

Applied to the T2 (a) spectrum, the ionized reflection model C provides a statistically significant improvement ($>3\sigma$) over the power-law one. The recovered photon index for the coronal emission is $\Gamma_{\rm corona} = 2.62^{+0.31}_{-0.39}$, with a high ionization parameter for the disk ($\log \xi \approx 3$). The latter is consistent with values commonly found in AGN with radiatively efficient accretion (D. R. Ballantyne et al. 2011), as expected for 5BZB J0630-2406 (Paper PI). While this model successfully describes the data, it presents certain challenges given that the inferred coronal photon index is toward softer values than the typical ones observed in Seyfert galaxies, where such spectral properties are well constrained.

### 6.4. Model D: Ionized Absorption from Jet–Cloud/Star Interaction

An alternative scenario for the spectral properties observed in T2 (b) is that the emission of the jet is absorbed by intervening material, such as a cloud or a star. To model it, we used the zxipcf absorption component, which is often used in obscured AGN. This model reproduces the effect of an ionized absorber, covering the source completely (i.e., covered fraction of 100%) and intercepting the primary continuum along the line of sight. We fixed the redshift of the absorber to that of the blazar. The results suggest a heavily obscured medium, with a high column density of $N_{\rm H,cloud} > 10^{24}\ {\rm cm^{-2}}$, and a well constrained ionization state of the cloud. Testing for partial covering absorption in this scenario indicates that it is consistent with fully absorbing the source along our line of sight.

Applying this scenario to epoch T2 (a), the best-fit model suggests a partially absorbed jet, with about 50%–65% of the emission absorbed, a dense interfering medium with $N_{\rm H,cloud} \approx 10^{24}\ {\rm cm^{-2}}$, along with a jet photon index of $\Gamma_{\rm jet} \sim 3.4$ and a high ionization state. Although the ionization parameter is not well constrained, this model provides a statistical improvement of $3.4\sigma$ compared to the power-law one.

### 6.5. Summary of Alternative Models

To assess the plausibility of the physical interpretations explored, we compared each alternative model to the null hypothesis in which the X-ray spectrum is entirely produced by emission from the jet. As shown in Table 5, all tested models show a statistically significant improvement over the null case above $2\sigma$.

While model A results in less statistical improvement than the other models, the X-ray spectrum can still be well described by a two-component power-law model across different epochs, e.g., T2 (a). Considering that this model relies on the fewest assumptions, it serves as a natural baseline among the tested scenarios. On the other hand, both models B and C offer greater statistical improvements. However, the increased number of free parameters indicates that the data are not sufficient to confidently constrain key quantities such as the spectral index of the coronal continuum or the reflection fraction. This limits our ability to robustly interpret these models despite their statistical advantage.

Finally, the scenario explored in model D provides the best fit to the data while successfully recovering the main physical parameters. Additionally, as explained in Section 6.4, the consistency of some recovered values with independent measurements in AGN makes this scenario noteworthy.

## 7. Multimessenger Implications and Potential Neutrino Connection

Following the identification of 5BZB J0630-2406 as a potential neutrino source, in our previous study we carried out a lepto-hadronic modeling of the quasi-simultaneous blazar SED focused on T2 (a). Within that model, the additional component in X-rays has been interpreted as the presence of a hadronic contribution in the jet, possibly from Bethe–Heitler processes or a mix of leptonic and hadronic components (Paper TI). In such environments, within regions of lower ambient matter density, neutrinos are expected to originate from $p\gamma$ interactions, leading to harder neutrino spectra at energies >100 TeV (e.g., TXS 0506+056; $dN/dE_\nu \propto E_\nu^{-2.0}$, C. D. Dermer et al. 2014; A. Reimer et al. 2019).

The spectral variability observed in X-rays and the presence of an additional X-ray spectral component across multiple observations of low jet activity, close in time, suggest changes in the underlying emission mechanisms during 2008–2015. This period coincides with the timeline during which IceCube observations revealed a neutrino hotspot from the direction of 5BZB J0630-2406, opening new interesting prospects in the potential blazar/neutrino connection. The previous section presented alternative explanations for the observed X-ray spectral properties. Among those explored, models B and C lead to coronal properties in tension with typically observed values. We further discuss implications of models A and D, which appear as plausible physical scenarios.

### 7.1. Implications for Coronal Component

The presence of a coronal emission component, as suggested by model A, prompts similarities to hadronic scenarios proposed for NGC 1068, a Seyfert galaxy proposed as a neutrino emitter (IceCube Collaboration et al. 2022). In such sources, which lack powerful jets, high-energy protons may be accelerated near the black hole (e.g., via stochastic acceleration or magnetic reconnection) and interact with the dense coronal photon field and surrounding matter near the black hole (K. Murase et al. 2020, 2016, 2024; A. Kheirandish et al. 2021; D. F. G. Fiorillo et al. 2024; D. Karavola et al. 2024). These interactions can produce detectable neutrino fluxes around the TeV range, characterized by a steep neutrino spectral index (for NGC 1068, a power-law fit to the observations shows $dN/dE_\nu \propto E_\nu^{-3.3}$). In such environments, $\gamma\gamma$ interactions dominate over $p\gamma$ processes as the dense photon field in the X-ray to soft $\gamma$-ray regime significantly increases the opacity for high-energy $\gamma$-rays through pair production. The high $\tau_{\gamma\gamma}$ optical depth suppresses the escape of $\gamma$-rays, effectively making $p\gamma$ interactions less efficient than *pp* processes. Consequently, these interactions predominantly emit radiation that is detectable primarily within the hard X-ray to soft $\gamma$-ray spectral range. Building on this, A. Neronov et al. (2024) propose a direct proportionality between the intrinsic hard X-ray luminosity and neutrino flux, particularly in Compton-thick sources, suggesting a linear scaling between these two messengers. This is less likely the case for blazars, where the X-ray band is dominated by the nonthermal emission of the powerful jet (see also E. Kun et al. 2024).

The transitional spectral properties observed in 5BZB J0630-2406 offer a unique opportunity to investigate the coronal component in a blazar and may serve as a bridge for understanding neutrino production mechanisms in both Seyfert galaxies and blazars. D. F. G. Fiorillo et al. (2025) revisited the corona hypothesis for neutrino production in TXS 0506+056, demonstrating that while magnetic reconnection can accelerate protons to tens of PeV, the resulting coronal neutrino emission remains insufficient to account for the observed IceCube spectrum. For TXS 0506+056, the estimated steady X-ray coronal luminosity ranges from $L_{\rm corona} = 4 \times 10^{43}$ erg s$^{-1}$ to $L_{\rm corona} = 4 \times 10^{44}$ erg s$^{-1}$. This is about an order of magnitude lower than what we find for 5BZB J0630-2406. Based on model A, the inferred coronal luminosity during the low-activity state reaches $L_{\rm corona} \sim 5 \times 10^{45}$ erg s$^{-1}$, while in the high-activity state, considering the measured 2–10 keV integrated flux as an upper bound, it could increase up to $L_{\rm corona} \sim 5 \times 10^{46}$ erg s$^{-1}$. As shown in recent studies (e.g., A. Neronov et al. 2025) we conclude that, for this blazar, the corona remains an interesting perspective to be explored as potential contributor to its neutrino emission.

### 7.2. Implications for Jet–Cloud/Star Interaction

In systems such as Seyfert galaxies, the general interpretation is that X-ray emission originates from the corona close to the SMBH, making variable clouds likely to interact with it. In blazars, any obscuration is more likely to occur on much larger spatial scales, involving the jet or external environments rather than the nuclear regions. The intervening material could be an ionized cloud, e.g., of the BLR or a star (W. Bednarek & R. J. Protheroe 1997; A. Dar & A. Laor 1997; A. T. Araudo et al. 2010, 2013). While the rate for such interactions to occur in blazars is expected to be some tens per month but remains unconstrained, the typical jet–cloud interaction timescales are of the order of hours (S. del Palacio et al. 2019). Therefore, one would not necessarily expect to observe the same phenomena across different epochs of years-long observations.

Jet–cloud/star interaction scenarios have been explored in studies on neutrino emission from blazars (e.g., K. Wang et al. 2022). Our derived gas column density for 5BZB J0630-2406 is consistent with the values ($N_{\rm H,cloud} \approx 10^{24}$ cm$^{-2}$) required for neutrino production in models proposing jet–BLR cloud interactions, as suggested for TXS 0506+056 (R.-Y. Liu et al. 2019). Such a scenario could potentially explain the neutrino hotspot associated with 5BZB J0630-2406.

## 8. Conclusion

In this study, we have conducted a comprehensive analysis of the X-ray emission properties of the blazar 5BZB J0630-2406, utilizing data from XMM-Newton, NuSTAR, Swift, and eROSITA across different epochs. Our analysis highlights the transitional X-ray properties of 5BZB J0630-2406, with key findings:

1. The X-ray spectra of 5BZB J0630-2406 show significant variability in the 2.0–10.0 keV band. During higher flux states, the spectra adhere to a power-law model consistent with nonthermal jet emissions. In contrast, lower flux states reveal an additional spectral component, suggesting the potential contribution from coronal emission, jet–cloud/star interaction or other hadronic components.
2. The spectral evolution is observed during the timeframe when an IceCube neutrino hotspot is associated with 5BZB J0630-2406 (2008–2015, Paper I). Possible physical scenarios suggest a correlation between X-ray variability and variability in neutrino emission. Thus, regardless of the underlying mechanism behind the additional component, the X-ray flare observed in 2011 may mark a phase of enhanced neutrino production, favoring the detection of this blazar during these early years of operation of IceCube.
3. Simulations were performed to estimate the exposure required to confirm or rule out the additional spectral component. We showed that joint observations with XMM-Newton (33 ks) and NuSTAR (80 ks) would reliably detect or rule out the additional component in the spectrum at high statistical significance (∼5$\sigma$). Such observations would provide crucial insights into whether the component is a transient phenomenon or an intrinsic property of the blazar.

4. The detection of spectral changes in the X-ray band in 5BZB J0630-2406 offers a new perspective for studying neutrino production in blazars, and possibly distinguishing between hadronic mechanisms. If the neutrino spectrum of 5BZB J0630-2406 follows a harder spectrum ($dN/dE_\nu \propto E_\nu^{-2}$), this would favor $p\gamma$ interactions within the jet, while a softer neutrino spectrum would suggest a common coronal origin as in the similar soft emission from NGC 1068.

*Facilities:* Swift, XMM, NuSTAR, eROSITA, IceCube.

*Software:* ASTROPY (Astropy Collaboration et al. 2013, 2018), TOPCAT (M. B. Taylor 2005).

## Acknowledgments

We thank the anonymous referees for their useful comments and suggestions that helped improve this manuscript. The authors thank D. Fiorillo, M. Boughelilba, and D. Prokhorov for the discussions related to this work. This work was supported by the European Research Council, ERC Starting grant MessMapp, S.B. Principal Investigator, under contract no. 949555. This work makes use of Swift and NuSTAR observations as part of proposal ID 2023182 and Swift ToO ID 20480, as well as public archival data. This research has made use of data from the NuSTAR mission, a project led by the California Institute of Technology, managed by the Jet Propulsion Laboratory, and funded by the National Aeronautics and Space Administration. Data analysis was performed using the NuSTAR Data Analysis Software (NuSTARDAS), jointly developed by the ASI Science Data Center (SSDC, Italy) and the California Institute of Technology (USA). We thank the NuSTAR Operations, Software, and Calibration teams for support with the execution and analysis of these observations. S.M. research is carried out with contribution of the Next Generation EU funds within the National Recovery and Resilience Plan (PNRR), Mission 4—Education and Research, Component 2—From Research to Business (M4C2), Investment Line 3.1—Strengthening and creation of Research Infrastructures, Project IR0000012—"CTA+—Cherenkov Telescope Array Plus." This work is based on data from eROSITA, the soft X-ray instrument aboard SRG, a joint Russian–German science mission supported by the Russian Space Agency (Roskosmos), in the interests of the Russian Academy of Sciences represented by its Space Research Institute (IKI), and the Deutsches Zentrum für Luft und Raumfahrt (DLR). The SRG spacecraft was built by Lavochkin Association (NPOL) and its subcontractors, and is operated by NPOL with support from the Max Planck Institute for Extraterrestrial Physics (MPE). The development and construction of the eROSITA X-ray instrument was led by MPE, with contributions from the Dr. Karl Remeis Observatory Bamberg & ECAP (FAU Erlangen-Nuernberg), the University of Hamburg Observatory, the Leibniz Institute for Astrophysics Potsdam (AIP), and the Institute for Astronomy and Astrophysics of the University of Tübingen, with the support of DLR and the Max Planck Society. The Argelander Institute for Astronomy of the University of Bonn and the Ludwig Maximilians Universität Munich also participated in the science preparation for eROSITA. This research has made use of data and/or software provided by the High Energy Astrophysics Science Archive Research Center (HEASARC), which is a service of the Astrophysics Science Division at NASA/GSFC. Part of this work is based on archival data, software, or online services provided by the ASI Space Science Data Center (SSDC). Based on observations obtained with XMM-Newton, an ESA science mission with instruments and contributions directly funded by ESA Member States and NASA.

## ORCID iDs

Jose Maria Sanchez Zaballa https://orcid.org/0009-0001-9486-1252
Sara Buson https://orcid.org/0000-0002-3308-324X
Stefano Marchesi https://orcid.org/0000-0001-5544-0749
Francesco Tombesi https://orcid.org/0000-0002-6562-8654
Thomas Dauser https://orcid.org/0000-0003-4583-9048
Joern Wilms https://orcid.org/0000-0003-2065-5410
Alessandra Azzollini https://orcid.org/0000-0002-2515-1353

## References

Ackermann, M., Ajello, M., An, H., et al. 2016, ApJ, 820, 72
Araudo, A. T., Bosch-Ramon, V., & Romero, G. E. 2010, A&A, 522, A97
Araudo, A. T., Bosch-Ramon, V., & Romero, G. E. 2013, MNRAS, 436, 3626
Arnaud, K. A. 1996, in ASP Conf. Ser. 101, Astronomical Data Analysis Software and Systems V, ed. G. H. Jacoby & J. Barnes (San Francisco, CA: ASP), 17
Astropy Collaboration, Price-Whelan, A. M., Sipöcz, B. M., et al. 2018, AJ, 156, 123
Astropy Collaboration, Robitaille, T. P., Tollerud, E. J., et al. 2013, A&A, 558, A33
Azzollini, A., Buson, S., Coleiro, A., et al. 2025, arXiv:2507.03613
Ballantyne, D. R., McDuffie, J. R., & Rusin, J. S. 2011, ApJ, 734, 112
Baloković, M., Harrison, F. A., Madejski, G., et al. 2020, ApJ, 905, 41
Bednarek, W., & Protheroe, R. J. 1997, MNRAS, 287, L9
Bregman, J. N. 1990, A&ARv, 2, 125
Brunner, H., Liu, T., Lamer, G., et al. 2022, A&A, 661, A1
Buson, S., Tramacere, A., Pfeiffer, L., et al. 2022a, ApJL, 933, L43
Buson, S., Tramacere, A., Pfeiffer, L., et al. 2022b, ApJL, 934, L38
Buson, S., Tramacere, A., Oswald, L., et al. 2023, arXiv:2305.11263
Cash, W. 1979, ApJ, 228, 939
Comastri, A., Fossati, G., Ghisellini, G., & Molendi, S. 1997, ApJ, 480, 534
Dar, A., & Laor, A. 1997, ApJL, 478, L5
Dauser, T., García, J., Walton, D. J., et al. 2016, A&A, 590, A76
del Palacio, S., Bosch-Ramon, V., & Romero, G. E. 2019, A&A, 623, A101
Dermer, C. D., Murase, K., & Inoue, Y. 2014, JHEAP, 3, 29
Fichet de Clairfontaine, G., Buson, S., Pfeiffer, L., et al. 2023, ApJL, 958, L2
Fiorillo, D. F. G., Petropoulou, M., Comisso, L., Peretti, E., & Sironi, L. 2024, ApJL, 961, L14
Fiorillo, D. F. G., Testagrossa, F., Petropoulou, M., & Winter, W. 2025, ApJ, 986, 104
García, J., Dauser, T., Reynolds, C. S., et al. 2013, ApJ, 768, 146
García, J., & Kallman, T. R. 2010, ApJ, 718, 695
García, J., Kallman, T. R., & Mushotzky, R. F. 2011, ApJ, 731, 131
Gehrels, N., Chincarini, G., Giommi, P., et al. 2004, ApJ, 611, 1005
Ghisellini, G., Tavecchio, F., Foschini, L., et al. 2012, MNRAS, 425, 1371
Grandi, P., & Palumbo, G. G. C. 2004, Sci, 306, 998
Harrison, F. A., Craig, W. W., Christensen, F. E., et al. 2013, ApJ, 770, 103
IceCube Collaboration, Aartsen, M. G., Ackermann, M., et al. 2018, Sci, 361, eaat1378
IceCube Collaboration, Abbasi, R., Ackermann, M., et al. 2022, Sci, 378, 538
Kalberla, P. M. W., Burton, W. B., Hartmann, D., et al. 2005, A&A, 440, 775
Karavola, D., Petropoulou, M., Fiorillo, D. F. G., Comisso, L., & Sironi, L. 2025, JCAP, 2025, 075
Kheirandish, A., Murase, K., & Kimura, S. S. 2021, ApJ, 922, 45
Kun, E., Bartos, I., Tjus, J. B., et al. 2024, PhRvD, 110, 123014

Lainez, M., Dominguez, A., Paliya, V. S., et al. 2023, ICRC (Nagoya), 38, 558, ICRC
Laurenti, M., Piconcelli, E., Zappacosta, L., et al. 2022, A&A, 657, A57
Liu, R.-Y., Wang, K., Xue, R., et al. 2019, PhRvD, 99, 063008
Madsen, K. K., Harrison, F. A., Markwardt, C. B., et al. 2015, ApJS, 220, 8
Mannheim, K. 1993, A&A, 269, 67
Marchesi, S., Iuliano, A., Prandini, E., et al. 2025, A&A, 693, A142
Marchesi, S., Zhao, X., Torres-Albà, N., et al. 2022, ApJ, 935, 114
Merloni, A., Lamer, G., Liu, T., et al. 2024, A&A, 682, A34
Miller, L., Turner, T. J., Reeves, J. N., et al. 2006, A&A, 453, L13
Murase, K., Guetta, D., & Ahlers, M. 2016, PhRvL, 116, 071101
Murase, K., Karwin, C. M., Kimura, S. S., Ajello, M., & Buson, S. 2024, ApJL, 961, L34
Murase, K., Kimura, S. S., & Mészáros, P. 2020, PhRvL, 125
Mutie, I. M., del Palacio, S., Beswick, R. J., et al. 2025, MNRAS, 539, 808
Nandra, K., O'Neill, P. M., George, I. M., & Reeves, J. N. 2007, MNRAS, 382, 194
Neronov, A., Kalashev, O., Semikoz, D. V., Savchenko, D., & Poleshchuk, M. 2025, arXiv:2503.16273
Neronov, A., Savchenko, D., & Semikoz, D. V. 2024, PhRvL, 132, 101002
Oikonomou, F., Petropoulou, M., Murase, K., et al. 2021, JCAP, 2021, 082
Padovani, P., Oikonomou, F., Petropoulou, M., Giommi, P., & Resconi, E. 2019, MNRAS Letters, 484, L104
Petropoulou, M., Dimitrakoudis, S., Padovani, P., Mastichiadis, A., & Resconi, E. 2015, MNRAS, 448, 2412
Petropoulou, M., Murase, K., Santander, M., et al. 2020, ApJ, 891, 115
Pizzetti, A., Torres-Albà, N., Marchesi, S., et al. 2022, ApJ, 936, 149
Predehl, P., Andritschke, R., Arefiev, V., et al. 2021, A&A, 647, A1
Reimer, A., Böttcher, M., & Buson, S. 2019, ApJ, 881, 46
Sanders, D. B., Phinney, E. S., Neugebauer, G., Soifer, B. T., & Matthews, K. 1989, ApJ, 347, 29
Shaw, M. S., Romani, R. W., Cotter, G., et al. 2013, ApJ, 764, 135
Strüder, L., Briel, U., Dennerl, K., et al. 2001, A&A, 365, L18
Sun, X.-N., Zhang, J., Lin, D.-B., et al. 2014, ApJ, 798, 43
Sunyaev, R., Arefiev, V., Babyshkin, V., et al. 2021, A&A, 656, A132
Swain, S., Stalin, C. S., Paliya, V. S., & Saikia, D. J. 2025, MNRAS, 537, 97
Tarter, C. B., Tucker, W. H., & Salpeter, E. E. 1969, ApJ, 156, 943
Taylor, M. B. 2005, in ASP Conf. Ser. 347, Astronomical Data Analysis Software and Systems XIV, ed. P. Shopbell, M. Britton, & R. Ebert (San Francisco, CA: ASP), 29
Tombesi, F., Meléndez, M., Veilleux, S., et al. 2015, Natur, 519, 436
Turner, M. J. L., Abbey, A., Arnaud, M., et al. 2001, A&A, 365, L27
Verner, D. A., Ferland, G. J., Korista, K. T., & Yakovlev, D. G. 1996, ApJ, 465, 487
Veronica, A., Reiprich, T., Pacaud, F., et al. 2025, A&A, 694, A168
Wang, K., Liu, R.-Y., Li, Z., Wang, X.-Y., & Dai, Z.-G. 2022, Univ, 9, 1
Wilms, J., Allen, A., & McCray, R. 2000, ApJ, 542, 914
Zhang, H., Fang, K., Li, H., et al. 2019, ApJ, 876, 109